\documentclass{article}
\usepackage[sumlimits]{amsmath}
\usepackage{amssymb}
\usepackage{amsthm}

\usepackage{color}

\def\twoheaddownarrow{\rlap{$\downarrow$}\raise-.5ex\hbox{$\downarrow$}}
\def\twoheaduparrow{\rlap{$\uparrow$}\raise.5ex\hbox{$\uparrow$}}

\usepackage[hidelinks]{hyperref}

\usepackage[a4paper,portrait,margin=2cm]{geometry}

\usepackage{graphicx}

\usepackage{stmaryrd} 

\usepackage{tikz}
\usetikzlibrary{arrows}
\usetikzlibrary{fadings}

\usepackage{tikz-cd}
\tikzcdset{every label/.append style = {font = \small}}

\usepackage{todonotes}

\newcommand{\hideEM}[1]{}
 
\usepackage{xspace}
\def\ie{i.e.\xspace}
\def\eg{e.g.\xspace} 

\newcommand{\id}{\mathsf{id}}
\newcommand{\Id}{\mathsf{Id}}

\DeclareFontFamily{OT1}{pzc}{}
\DeclareFontShape{OT1}{pzc}{m}{it}{<->s*[1.30]pzcmi7t}{}
\DeclareMathAlphabet{\mathpzc}{OT1}{pzc}{m}{it}
\def\ct#1{\mathpzc{#1}} 

\newcommand{\Alg}{\mathsf{A}}

\newcommand{\RQ}{\mathbb{R}}
\newcommand{\NQ}{\mathbb{N}}
\newcommand{\BQ}{\Sigma}

\newcommand{\qle}{\sqsubseteq} 
\newcommand{\qlt}{\sqsubset} 
\newcommand{\qJ}{\bigsqcup} 
\newcommand{\qj}{\sqcup} 
\newcommand{\qM}{\bigsqcap}
\newcommand{\qm}{\sqcap}
\newcommand{\qT}{\otimes} 
\newcommand{\qt}{\top} 
\newcommand{\qb}{\bot} 
\newcommand{\qI}{\mathsf{u}} 
\newcommand{\qlr}{\backslash} 
\newcommand{\qrr}{/} 

\newcommand{\qwa}{\twoheaduparrow} 
\newcommand{\qwb}{\twoheaddownarrow}
\newcommand{\qLowerSet}{\mathop{\downarrow}}  

\newcommand{\KE}{\mathsf{K}} 
\newcommand{\PE}{\mathsf{J}} 

\newcommand{\V}{\ct{V}}
\newcommand{\A}{\ct{A}}
\newcommand{\Set}{\ct{Set}}
\newcommand{\CL}{\ct{CL}}
\newcommand{\PCL}{\ct{PCL}}
\newcommand{\Po}{\ct{Po}}
\newcommand{\Pos}{\Po_0}
\newcommand{\MS}{\ct{Met}}
\newcommand{\Quan}{\ct{Qnt}} 
\newcommand{\Top}{\ct{Top}}

\newcommand{\s}[1]{{#1}_s}

\newcommand{\Mon}{\mathsf{Mon}}  

\newcommand{\R}{R}
\renewcommand{\r}{r}
\newcommand{\sr}{s}
\newcommand{\T}{\mathsf{T}}
\newcommand{\inT}{\mathsf{in}} 

\newcommand{\cl}[1]{\overline{#1}}
\newcommand{\inv}[1]{{#1}^{-1}}

\newcommand{\CS}{\mathsf{C}}
\newcommand{\setf}[1]{\ensuremath{\{{#1}\}}}
\newcommand{\setb}[2]{\ensuremath{\{{#1}\mid{#2}\}}}
\newcommand{\famb}[2]{\ensuremath{({#1}\mid{#2})}}
\newcommand{\pair}[1]{\ensuremath{\langle{#1}\rangle}}
\newcommand{\DM}{\mathsf{D}}
\newcommand{\CM}{\mathsf{C}}
\newcommand{\PM}{\mathsf{P}}

\newcommand{\defeq}{\stackrel{\vartriangle}{=}}
\newcommand{\defiff}{\stackrel{\vartriangle}{\iff}}

\newcommand{\eqc}[2][]{[#2]_{#1}}

\newcommand{\conv}{\odot} 
\newcommand{\Star}{\ensuremath{(\star)}\xspace}
\newcommand{\SStar}{\ensuremath{(\star\star)}\xspace}

\theoremstyle{plain} 

\newtheorem{definition}{Definition}[section]

\newtheorem{corollary}[definition]{Corollary}
\newtheorem{lemma}[definition]{Lemma}

\newtheorem{proposition}[definition]{Proposition}

\newtheorem{theorem}[definition]{Theorem}

\theoremstyle{definition}

\newtheorem{example}[definition]{Example}
\newtheorem{remark}[definition]{Remark}

\begin{document}

\title{\textbf{Robust Topology and the Hausdorff-Smyth Monad\\
  on Metric Spaces over Continuous Quantales}}

\author{Francesco Dagnino\thanks{DIBRIS, University of Genoa,
    Italy, Email:\href{mailto:francesco.dagnino@unige.it}{\texttt{francesco.dagnino@unige.it}}, ORCID: 0000-0003-3599-3535}
    \and
    Amin Farjudian\thanks{School of
      Mathematics, University of Birmingham, United Kingdom, Email:
    \href{mailto:A.Farjudian@bham.ac.uk}
    {\texttt{A.Farjudian@bham.ac.uk}}, ORCID: 0000-0002-1879-0763, }
  \and
Eugenio Moggi\thanks{DIBRIS, University of Genoa,
    Italy, Email:\href{mailto:moggi@unige.it}{\texttt{moggi@unige.it}}, ORCID: 0000-0001-8018-6543}
  }

\date{}
  
\maketitle

\begin{abstract}
We define a (preorder-enriched) category $\MS$ of
  quantale-valued metric spaces and uniformly continuous maps, with
  the essential requirement that the quantales are continuous.  For
  each object $(X,d,Q)$ in this category, where $X$ is the carrier
  set, $Q$ is a continuous quantale, and $d: X \times X \to Q$ is the
  metric, we consider a topology $\tau_d$ on $X$, which generalizes
  the open ball topology, and a topology $\tau_{d,R}$ on the powerset
  $\PM(X)$, called the robust topology, which captures robustness with
  respect to small perturbations of parameters. We define a
  (preorder-enriched) monad $\PM_S$ on $\MS$, called the
  Hausdorff-Smyth monad, which captures the robust topology, in the
  sense that the open ball topology of the object $\PM_S(X,d,Q)$
  coincides with the robust topology $\tau_{d,R}$ for the object
  $(X,d,Q)$. We prove that every topology arises from a
  quantale-valued metric. As such, our framework provides a foundation
  for quantitative reasoning about imprecision and robustness in a
  wide range of computational and physical systems.   
\end{abstract}


\section{Introduction}

Robustness is a central topic in analysis of systems, including
cyber-physical and machine learning systems. Robustness is a broad
term with various instances, and studying each instance requires a
suitable mathematical theory.  We focus on the specific case of
robustness with respect to small perturbations of parameters.

\paragraph{Robust analyses.} The notions of imprecision and robustness
are relevant in the context of software tools for analysis of
hybrid/continuous systems.  These tools manipulate (formal
descriptions of) mathematical models, providing a simplified
description of the system and its environment.  A desirable
requirement is that simplifications should be \emph{safe}, \ie, if the
analysis implies that the model satisfies a property, then the system
also satisfies that property. Safe simplification is commonly achieved
by injecting \emph{non-determinism} in the model.  Non-determinism is
also useful in modeling \emph{known unknowns} in the environment and
features of the system about which we \emph{`do not care'}.
Another issue that arises in hybrid/continuous systems is the
\emph{imprecision} in observations. In fact, predictions based on a
mathematical model and observations on a real system can be compared
only up to the precision of measurements on the real system.
We say that \textbf{an analysis is robust} when it can cope with
\emph{small amounts} of imprecision in the model, \ie, if a robust
analysis implies that a model~$M$ has a property, then this property
is implied also for models that have a bit more non-determinism
than~$M$.
Working with metric spaces makes it possible to define imprecision
formally and to quantify the amount of non-determinism added to a
model.

Following~\cite{MoggiFT-ICTCS-2019} one
can identify analyses with monotonic maps between complete lattices
$\PM(X)$ of subsets of $X$ ordered by reverse inclusion\footnote{The
category of complete lattices and monotonic maps is the framework
proposed in~\cite{cousot1992abstract} for abstract interpretations.},
where $(X,d)$ is a metric space.
Since two subsets of $X$ with the same closure with respect to the
open ball topology induced by the metric $d$ are
\emph{indistinguishable}, even when imprecision is made arbitrarily
small, one can restrict analyses to complete lattices $\CS(X)$ of closed
subsets of metric spaces $(X,d)$.
In this setting robust analyses correspond to (monotonic) maps that
are continuous with respect to the \emph{robust topology} on $\CM(X)$
(see~\cite[Def.~A.1]{Moggi_Farjudian_Duracz_Taha:Reachability_Hybrid:2018}).

\paragraph{Generalized metric spaces.}  In~\cite{lawvere1973metric},
Lawvere proposed viewing metric spaces as small categories enriched
over the monoidal category $\RQ_+$, whose objects are the extended
non-negative real numbers, there is an arrow $x\to y$ if and only if
$x\geq y$, and $+$ and $0$ provide the monoidal structure.  In this
way, one recovers several notions and results for metric spaces as
instances of those for enriched categories~\cite{kelly1982basic}.

The monoidal category $\RQ_+$ used by Lawvere belongs to the class of
small (co)complete posetal categories, where the tensor commutes with
colimits.
These categories are called
\emph{quantales}~\cite{Mulvey:Second_topology_conference:1986,Niefield_Rosenthal:quantales:1988,AbramskyV93}
and small categories enriched over a quantale $Q$ are dubbed
\emph{$Q$-metric spaces}.
Quantale-valued metric spaces are increasingly used for
\emph{quantitative reasoning} on programs/systems, and for defining
various notions of \emph{behavioral metrics}
\cite{Gavazzo18,BonchiKP18,DalLagoGY19,Pistone21,SprungerKDH21,GavazzoF23,DahlqvistN23}.
The use of quantitative methods is important in coping with the
uncertainty/imprecision that arises in the analysis of, {\eg},
probabilistic programs or systems interacting with physical processes.
In these contexts, quantales provide a flexible framework which allows
choosing the most suitable notion of metric for the specific analysis
one is interested in.
Quantales arise implicitly also in the analysis of algorithms, where
\emph{costs} are values in certain quantales (see
Example~\ref{ex:cost-quantale}).

In the context of fuzzy systems, quantales offer a powerful and
abstract algebraic structure that can model the graded, non-binary
nature of fuzzy set theory. They serve as a bridge between fuzzy
logic, lattice theory, and category theory, enabling deeper insights
and
generalizations~\cite{Hohle_Kubiak:non_commutative_quantale:2011,Tao_Lai_Zhang:Quantale_valued_preorders:2014,GutierrezGarcia_Hohle_Kubiak:Tensor_Quantales:2017}. They
also yield an expressive algebraic structure for fuzzy sets of higher
types~\cite{Eklund_Kortelainen_Lofstrand:Quantales_Higher_Types:2025}.

Quantales provide a useful compromise between arbitrary monoidal
categories and the specific instance
$\RQ_+$~\cite{FlaggK97,HofmannST14,CookW21}.
Beside a substantial simplification of the theory, restricting to
quantales enables the use of well-known order-theoretic notions, which
do not have immediate counterparts in arbitrary monoidal categories,
but are crucial for relating $Q$-metric spaces to other structures
such as topological spaces.
It is therefore appropriate to ask whether the notions of imprecision,
robustness, and robust topology can be extended to $Q$-metric spaces.

\begin{itemize}
\item In~\cite[Remark
  2.26]{Farjudian_Moggi:Robustness_Scott_Continuity_Computability:2023}
  we observed that for an ordinary metric space $(X,d)$ the robust topology on
  $\CM(X)$ is the open ball topology for a certain $\RQ_+$-metric,
  called \emph{Hausdorff-Smyth hemi-metric}\footnote{The name
  `Hausdorff-Smyth' is used in~\cite{goubault2008simulation} for an
  asymmetric variant of the Hausdorff metric related to the upper
  Vietoris topology on the Smyth powerdomain of a topological space,
  while 'hemi-metric' is a synonym for $\RQ_+$-metric, see also
 ~\cite{Goubault-Larrecq:Non_Hausdorff_topology:2013}.}
  in~\cite{goubault2008simulation}.

\item In~\cite{DFM:ICTAC:2023} we showed that the notions of
  imprecision, open ball topology $\tau_d$ on $X$, and robust topology
  $\tau_{d,R}$ on $\PM(X)$ generalize to any $Q$-metric space $(X,d)$,
  provided $Q$ is a continuous quantale. We also showed that the
  Hausdorff-Smyth hemi-metric generalizes to $Q$-metric spaces. More
  precisely, it is part of a (preorder-enriched) monad on the category
  of $Q$-metric spaces and short maps, which maps a $Q$-metric on $X$
  to a $Q$-metric on $\PM(X)$.
\end{itemize}
In~\cite{DFM:ICTAC:2023} we proved that the Hausdorff-Smyth $Q$-metric on
$\PM(X)$ captures the robust topology $\tau_{d,R}$, but only when $Q$
is a linear and non-trivial quantale.

\paragraph{Contributions.}

Some of the material in this article has appeared previously in the
proceedings of the International Colloquium on Theoretical Aspects of
Computing (ICTAC 2023)~\cite{DFM:ICTAC:2023}, where, due to page
limit, the proofs had to be omitted. In the current article, we
present the missing proofs.

The main contribution of the current article is replacing
(preorder-enriched) categories $\MS_Q$ of $Q$-metric spaces $(X,d)$
and short maps where $Q$ is a \emph{fixed} continuous quantale ({\ie},
the framework of~\cite{DFM:ICTAC:2023}), with one (preorder-enriched)
category $\MS$ of quantale-valued metric spaces $(X,d,Q)$ and
uniformly continuous maps, where different objects may use different
continuous quantales.
By allowing the quantales to vary, we can define on $\MS$ a
(preorder-enriched) monad $\PM_S$, also called Hausdorff-Smyth, which
captures the robust topology (see
Theorem~\ref{theorem:robust_top_induced_by_Hausdorff}).
In Example~\ref{example:coutnerexample_non_linear_quantale} we show
that for a (fixed) finite non-linear quantale $Q$ and a finite
$Q$-metric $(X,d)$, no $Q$-metric on $\PM(X)$ can capture the robust
topology $\tau_{d,R}$. As such, allowing the quantales to vary is
essential for coping with such examples.

\paragraph{Structure of the Paper.}
We summarize the structure of the paper, and highlight how each
section relates to~\cite{DFM:ICTAC:2023}. We point out that the
current manuscript is self-contained, and familiarity
with~\cite{DFM:ICTAC:2023} is not required for understanding the
content of the current article:

\begin{itemize}
\item Section~\ref{sec:maths} introduces some basic notation and
  terminology, and recalls some notions used in the rest of the paper,
  such as preorder-enriched categories and monads.

\item Section~\ref{sec:quantales} recalls the definition of quantale
  and continuous quantale (and considers also refinements of
  continuous quantales), and defines the preorder-enriched category
  $\Quan$ of continuous quantales and \emph{lax-unital} Scott
  continuous maps.
  The definition of $\Quan$ differs from that in
 ~\cite{DFM:ICTAC:2023}, since here we require the quantales to be
  continuous and the arrows to be Scott continuous.  These
  requirements do not represent a significant restriction, since
  Propositions~\ref{prop-D-construction}~and~\ref{prop-D-alg} say that
  any ordered monoid $P$ (thus also a quantale) can be embedded into a
  continuous quantale $\DM(P)$ in such a way that every monotonic map
  $f:P_1\to P_2$ between ordered monoids extends to a join-preserving
  (thus Scott continuous) map $\DM(f):\DM(P_1)\to\DM(P_2)$.

\item Section~\ref{sec:QMS} defines the categories of quantale-valued
  metric spaces $\MS_\A$, where $\A$ is a sub-category of $\Quan$.
  In particular, $\MS_\Quan$, denoted $\MS$ for short, can be
  described as the category of quantale-values metric spaces and
  uniformly continuous maps, while $\MS_Q$ (with $Q$ continuous
  quantale) is the category of $Q$-valued metric spaces and short
  maps.
  In this paper we focus on $\MS$, while in~\cite{DFM:ICTAC:2023} the
  main focus is on $\MS_Q$ and $\MS$ (denoted $\MS_\qI$) is only
  mentioned en passant.

  The definitions of open ball topology $\tau_d$, dual open ball
  topology $\tau^o_d$, and robust topology $\tau_{d,R}$ for an object
  $(X,d,Q)$ in $\MS$ are taken from~\cite{DFM:ICTAC:2023}. In
  Section~\ref{subsec:Extensive_Metrization}, we prove an extensive
  metrization result (Theorem~\ref{thm:extensive_metrizability}) which
  shows that every topology coincides with the open ball topology
  $\tau_d$ for an object $(X,d,Q)$ in $\MS$.

  In Example~\ref{ex:tauR_counterexample}, we show that, while the
  robust topology $\tau_{d,R}$ is determined by the $Q$-metric $d$, in
  general, it is not determined by the corresponding open ball
  topology $\tau_d$.

\item Section~\ref{sec:Po-enriched}, like~\cite[Section
  5]{DFM:ICTAC:2023}, defines the full sub-category $\s\A$ of
  separated objects in a preorder-enriched category $\A$ and shows
  that any preorder-enriched monad on $\A$ can be transformed (in an
  optimal way) into one that factors through $\s\A$.

\item Section~\ref{sec:Monads} recalls the definition given in
 ~\cite{DFM:ICTAC:2023} of the Hausdorff-Smyth monad $\PM_Q$ on
  $\MS_Q$ (see Definition~\ref{def:PM_Q}), then defines the
  Hausdorff-Smyth monad $\PM_S$ on $\MS$ (see
  Definition~\ref{def:PM_S}), and finally proves that $\PM_S$ captures
  the robust topology (see
  Theorem~\ref{theorem:robust_top_induced_by_Hausdorff}).
  The main difference between the two monads is that $\PM_S$ involves
  a change of quantale mediated by a monad $\CM_S$ on $\Quan$ (see
  Definition~\ref{def:CM_S}).
\end{itemize}

\section{Mathematical Preliminaries}
\label{sec:maths}

We use the symbol `$\in$' for set membership ({\eg}, $x \in X$), but
we use `$:$' for membership of function types (\eg, $f: X \to Y$) and
to denote objects and arrows in categories (\eg, $X:\Top$ and
$f:\Top(X,Y)$).

The powerset of a set $X$ is denoted by $\PM(X)$. Subset inclusion is
denoted by $\subseteq$, whereas strict (proper) subset inclusion is
denoted by $\subset$.  The finite powerset ({\ie}, the set of finite
subsets) of $X$ is denoted by $\PM_f(X)$, and $A\subseteq_f B$ denotes
that $A$ is a finite subset of $B$.
We denote with $\omega$ the set of natural numbers, and identify a
natural number with the set of its predecessors, {\ie}, $0 =
\emptyset$ and $n+1 = \setf{0, \ldots, n}$, for any $n\in\omega$.
We assume some basic familiarity with order
theory~\cite{Goubault-Larrecq:Non_Hausdorff_topology:2013}. For a
subset $S$ of a partial order $(Q,\qle)$, we write $\qJ S$ for the
join (aka lub) of $S$ and $\qM S$ for the meet (aka glb) of $S$,
whenever they exist.  We write $x\qj y$ and $x \qm y$ for the binary
join and binary meet of two elements $x$ and $y$ of $Q$, respectively.
$\qb$ and $\qt$ denote the bottom and top elements of $Q$,
respectively.
We also assume basic familiarity with category
theory~\cite{borceux1994handbook}. In this article:
\begin{itemize}
\item $\Set$ denotes the category of sets and functions (alias maps).
\item $\Po$ denotes the category of preorders and monotonic maps, and
  $\Pos$ denotes its full (reflective) sub-category consisting of
  posets.
\item $\Top$ denotes the category of topological spaces and continuous
  maps, and $\Top_0$ denotes its full (reflective) sub-category
  consisting of $T_0$-spaces.
\end{itemize}
All the categories above have small limits and colimits. The
categories $\Set$,$\Po$ and $\Pos$ have also exponentials, thus
they are examples of \emph{symmetric monoidal closed}
categories~\cite{kelly1982basic}.
$\Set$, $\Po$ and $\Top$ are related by \emph{faithful forgetful
functors}, denoted $U$ in Fig.~\ref{fig:forget}, each having a
\emph{full\&faithful} left-adjoint $I$, thus $\Set$ and $\Po$ can be
viewed as a full sub-category of $\Po$ and $\Top$, respectively.
\begin{figure}[tb]
  \begin{equation*}
      \begin{tikzcd}[row sep = huge, column sep = 6em]
    \Set \arrow[r, hookrightarrow, "I"', yshift = -1.3ex] & \Po
    \arrow[l, "U"', "\top", yshift = 1.3ex]  \arrow[r, hookrightarrow,
    "I"', yshift = -1.3ex]& \Top \arrow[l, "U"', "\top", yshift = 1.3ex]
    \end{tikzcd}
  \end{equation*}
  All functors above are the identity on the hom-sets, while on
  objects they act as follows:
  \begin{itemize}
  \item $U:\Top\to\Po$ maps a topological space $(X,\tau)$ to the
    specialization preorder $(X,\leq_\tau)$, its left-adjoint $I$ maps a
    preorder $(X,\leq)$ to the topological space $(X,\tau_\leq)$,
    where $\tau_\leq$ is Alexandrov topology, thus $U\circ I$ is the
    identity functor on $\Po$.
  \item $U:\Po\to\Set$ maps a preorder $(X,\leq)$ to the set $X$,
    its left-adjoint $I$ maps a set $X$ to the discrete preorder
    $(X,=_X)$, thus $U\circ I$ is the identity functor on $\Set$.
  \end{itemize}
  \caption{Relations among $\Set$, $\Po$, and $\Top$.}
  \label{fig:forget}
\end{figure}

\subsection{Preorder-enriched Categories}

Every (locally small) category $\A$ can be viewed as a
\emph{$\Po$-enriched} category~\cite{kelly1982basic}, \ie, each
hom-set $\A(X,Y)$ can be equipped with a preorder, so that composition
is monotonic.  In general, there are several $\Po$-enrichments of a
category $\A$, the extreme cases are the trivial enrichment (\ie,
$f\leq g$ for every $f,g:\A(X;Y)$) and the discrete one (\ie, $f\leq
g\iff f=g$).
The \emph{standard} $\Po$-enrichment of $\Po$ is the preorder on
$\Po(X,Y)$ given by the pointwise preorder (on the set of monotonic
maps from $X$ to $Y$), \ie, the exponential $Y^X$ in $\Po$.

An ordinary functor $F:\A\to\A'$ from a category $\A$ to a
$\Po$-enriched category $\A'$ allows to make $\A$ and $F$
$\Po$-enriched by taking as preorder on $\A(X,Y)$ the biggest one
making $F:\A(X,Y)\to\A'(FX,FY)$ monotonic, \ie, $f\leq g\defiff
Ff\leq' Fg$.
The \emph{standard} $\Po$-enrichment of $\Set$ is the discrete
preorder (induced by the functor $I:\Set\to\Po$), while that
of $\Top$ is the one induced by the functor $U:\Top\to\Po$.
\begin{example}
All functors in Fig.~\ref{fig:forget}, except $U:\Po\to\Set$, are
$\Po$-enriched with respect to the standard $\Po$-enrichments of
$\Set$, $\Po$ and $\Top$.  If one wants to make $U:\Po\to\Set$
$\Po$-enriched, then one must take the trivial $\Po$-enrichment of
$\Set$ (or the discrete $\Po$-enrichment of $\Po$).
\end{example}
$\Po$-enriched categories are a degenerate case of 2-categories, where
the notions of adjunction and equivalence (see~\cite{KellyStreet74})
specialize as follows.
\begin{definition}\label{def:po-adj} Given a pair of maps
  \begin{tikzcd}[column sep = large]
    X \arrow[r, "f" description , yshift =
    1.1ex] & Y  \arrow[l, "g" description , yshift = -1.1ex]
  \end{tikzcd}in a $\Po$-enriched category $\A$
\begin{enumerate}
\item $(f,g)$ is an \textbf{adjunction} (notation $f\dashv g$) $\defiff$
  $f\circ g\leq\id_Y$ and $\id_X\leq g\circ f$, in which $f$ and $g$
  are called left- and right-adjoint, respectively.
\item $(f,g)$ is an \textbf{equivalence} $\defiff$ $\id_Y\leq f\circ
  g\leq\id_Y$ and $\id_X\leq g\circ f\leq \id_X$.
\end{enumerate}
\end{definition}

In enriched Category Theory (see~\cite{kelly1982basic}) the definition
of full\&faithful functor requires some care.
\begin{definition}\label{def:po-ff}
  A $\Po$-enriched functor $F:\A\to\A'$ is \textbf{full\&faithful},
  notation $F:\A\hookrightarrow\A'$, when the maps
  $F_{X,Y}:\A(X,Y)\to\A'(FX,FY)$ are iso in $\Po$.
  Moreover, a $\Po$-enriched sub-category $\A$ of $\A'$ is
  \textbf{full}, when the $\Po$-enriched inclusion functor is
  full\&faithful.
\end{definition}
\begin{example}
The $\Po$-enriched functors $I:\Set\to\Po$ and $I:\Po\to\Top$ in
Fig.~\ref{fig:forget} are full\&faithful, when the categories involved
have the standard $\Po$-enrichments.
\end{example}

\subsection{Preorder-enriched Monads}

Since $\Po$-enriched categories form a 2-category, a $\Po$-enriched
monad is a monad in this 2-category (see~\cite{KellyStreet74}).
However, we adopt a simpler equivalent presentation in terms of
Kleisli triple (see~\cite{Manes1976,Moggi:notions_monads:1991}).
Monads and $\Pos$-enriched monads are instances of $\Po$-enriched
monads, since $\Set$ and $\Pos$ are sub-categories of $\Po$.

\begin{definition}\label{def:po-monad}
  A $\Po$-enriched monad on $\A$ is a triple $\hat{M}=(M,\eta,-^*)$,
  where:
 \begin{itemize}
 \item $M$ is a function on the objects of $\A$
 \item $\eta$ is a family of arrows $\eta_X:\A(X,MX)$ with $X:\A$
 \item $-^*$ is a family of monotonic maps $\A(X,MY)\to\A(MX,MY)$
   with $X,Y:\A$
 \end{itemize}
 satisfying the following equations:
 \begin{equation}\label{eq:monad}
   f^*\circ\eta_X=f\quad,\quad
   \eta_X^*=\id_{MX}\quad,\quad
   g^*\circ f^*=(g^*\circ f)^*.
 \end{equation}
Given two $\Po$-enriched monads $\hat{M}=(M,\eta,-^*)$ on $\A$ and
$\hat{M}'=(M',\eta',-^{*'})$ on $\A'$, a monad map from $\hat{M}$
to $\hat{M}'$ is a pair $(U,\theta)$, where:
 \begin{itemize}
 \item $U:\A\to\A'$ is a $\Po$-enriched functor
 \item $\theta$ is a family of maps $\theta_X:\A'(U(MX),M'(UX))$ with
   $X:\A$
 \end{itemize}
 satisfying the following equations:
  \begin{equation}\label{eq:monad:gmap}
    \theta_X\circ U(\eta_X)=\eta'_{UX}\quad,\quad
    \theta_Y\circ U(f^*)=(\theta_Y\circ Uf)^{*'}\circ\theta_X .
  \end{equation}
\end{definition}
The bijective correspondence between monads $(M,\eta,\mu)$ on $\A$,
where $M$ is a endofunctor on $\A$ and $\eta$ and $\mu$ are natural
transformations, and Kleisli triples $(M,\eta,-^*)$ extends verbatim
to the $\Po$-enriched case, namely:
\begin{itemize}
\item if $f:A(X,Y)$, then $Mf\defeq(\eta_Y\circ f)^*:\A(MX,MY)$
\item $\mu_X\defeq\id_{MX}^*:\A(M^2X,MX)$, and
\item if $f:\A(X,MY)$, then $f^*=\mu_Y\circ Mf:\A(MX,MY)$.
\end{itemize}
In the rest of the paper we do not use different superscripts to
distinguish the units $\eta$ and Kleisli extensions $f^*$ for
different monads, since it is usually clear from the context.

$\Po$-enriched monads and monad maps between them form a category
(more precisely a 2-category).  However, we are mainly interested in
$\Po$-enriched monads on a specific $\Po$-enriched category $\A$,
therefore we take a restricted notion on monad map, where $U$ is the
identity functor on $\A$.
\begin{definition}\label{def:Mon}
  The category $\Mon(\A)$ of $\Po$-enriched monads on $\A$ is
  \begin{description}
  \item[objects:] $\Po$-enriched monads $\hat{M}=(M,\eta,-^*)$ on $\A$
  \item[arrows:] $\theta$ from $\hat{M}$ to $\hat{M}'$ are families of
    maps such that $(\Id_\A,\theta)$ is a monad map from $\hat{M}$ to
    $\hat{M}'$, \ie,
    \begin{equation}\label{eq:monad:map}
      \theta_X\circ \eta_X=\eta'_X\quad,\quad
      \theta_Y\circ f^*=(\theta_Y\circ f)^{*'}\circ\theta_X .
    \end{equation}
  \end{description}
\end{definition}
Another notion related to monad maps is lifting of a $\Po$-enriched monad
along a $\Po$-enriched functor.
\begin{definition}\label{def:monad:lift}
  A $\Po$-enriched monad $\hat{M}$ on $\A$ is a \textbf{lifting} of
  the $\Po$-enriched monad $\hat{M'}$ on $\A'$ along the $\Po$-enriched
  functor $U:\A\to\A'$ $\defiff$ $(U,\id)$ is a monad map from
  $\hat{M}$ to $\hat{M}'$, where $\id$ is the identity natural
  transformation on the functor $U\circ M:\A\to\A'$,  \ie,
  \begin{equation}\label{eq:monad:lift}
    U(MX)=M'(UX)\quad,\quad
      U(\eta_X)=\eta'_X\quad,\quad
      U(f^*)=(Uf)^{*'} .
    \end{equation}
\end{definition}
\begin{remark}
  If $\A$ is a sub-category of $\A'$, then there is at most one
  lifting $\hat{M}$ of $\hat{M'}$ along the inclusion $\A\hookrightarrow\A'$.
  In this case, it is more appropriate to call $\hat{M}$ the
  \emph{restriction} of $\hat{M'}$ to $\A$.

  If $U:\A\to\A'$ is faithful, then there can be several lifting of
  $\hat{M'}$ along $U$.  However, if $\hat{M}_1$ and $\hat{M}_2$ are
  two liftings of $\hat{M}'$ such that $M_1=M_2$, then also the other
  components of the triples $\hat{M}_1$ and $\hat{M}_2$ are equal.
\end{remark}

\section{Quantales}
\label{sec:quantales}

Quantales~\cite{Mulvey:Second_topology_conference:1986,Niefield_Rosenthal:quantales:1988,AbramskyV93}
are degenerate instances of monoidal
categories~\cite{kelly1982basic}.
Fig.~\ref{fig:cat-poset} summarizes the correspondence between
category-theoretic notions and their order-theoretic counterparts.
In particular, posets correspond to categories that are \emph{small,
posetal and skeletal}\footnote{A category is small when its arrows
form a set, posetal when between two objects there is at most one
arrow, and skeletal when isomorphic objects are equal.}; and complete
lattices correspond to categories that are small, skeletal and
\emph{complete}\footnote{A small category which is also \emph{small
complete} is necessarily posetal, complete and co-complete.}.
\begin{figure}[tb]
  \begin{center}
  \begin{tabular}{|c|c|}
    \hline
    category $\A$&
    poset $P$
    \\\hline
    functor $F:\A\to\A'$&
    monotonic map $f:P\to P'$
    \\\hline 
    natural transformation $\sigma:F\to^.F'$&
    pointwise order $f\leq f'$
    \\\hline
    (co-)complete category&
    complete lattice
    \\\hline
    monoidal category&
    ordered monoid
    \\\hline
    (co-)complete bi-closed monoidal category&
    quantale
    \\\hline
  \end{tabular}
  \end{center}
  Posets correspond to categories that are small, posetal and skeletal.\\
  Complete lattices correspond to categories that are small, skeletal
  and (co-)complete.
  \caption{From categories to posets.}
  \label{fig:cat-poset}
\end{figure}

\begin{definition}[Quantale]
  \label{def:quantale}
  A \textbf{quantale} $(Q,\qle,\qT)$ is a complete lattice $(Q,\qle)$
  with a monoid structure $(Q,\qT,\qI)$, where $\qT$ and $\qI$ are
  called \textbf{tensor} and \textbf{unit}, respectively, and
  the following distributivity laws hold:
  \begin{equation}
    \label{eq:quantale_dist_laws}
    \forall x \in Q.\ \forall S \in \PM(Q):\quad 
    \left\{
      \begin{array}{rcl}
        x\qT (\qJ S) & = & \qJ \setb{x\qT y}{y\in S},\\
        (\qJ S)\qT x & =&  \qJ \setb{y\qT x}{y\in S}.\\
      \end{array}
    \right.        
  \end{equation}
  A quantale is \textbf{finite} when $Q$ is finite, \textbf{trivial}
  when $\bot=\qI$, \textbf{affine} when $\qI = \qt$, \textbf{linear}
  when $\qle$ is a linear order, and \textbf{commutative} when $\qT$
  is commutative, in which case the two distributivity laws
  in~\eqref{eq:quantale_dist_laws} are inter-derivable.
  A \textbf{locale}\footnote{Alternative names for locale are frame
  and complete Heyting algebra, see~\cite{Johnstone86}.} is a quantale
  where $\qT$ is binary meet $\qm$, in which case, it is necessarily
  commutative and affine.
\end{definition}
The distributivity laws imply that:
\begin{itemize}
\item The operator $\qT$ is monotonic. Thus, $(Q,\qT,\qI)$ makes
  $(Q,\qle)$ a (strict) monoidal category.
\item The operator $\qT$ (viewed as a functor) preserves colimits, in
  particular $\qb\qT x=\qb=x\qT\qb$.
\end{itemize}
These properties imply that the functors $x\qT -$ and $-\qT y$ have
right-adjoints $x\qlr -$ and $-\qrr y$, \ie,
\begin{equation*}
  \forall x, y, z \in Q:\quad x\qT y\qle z \iff y\qle x\qlr z \iff
  x\qle z\qrr y
\end{equation*}
given by $x\qlr z = \qJ\setb{y}{x\qT y\qle z}$ and $z\qrr y =
\qJ\setb{x}{x\qT y\qle z}$, and called left- and right-residual,
respectively.
In commutative quantales the two residuals coincide (\ie, $x\qlr
z=z\qrr x$) and are usually denoted $[x,z]$.

\begin{example}\label{ex:quantale}
  We give some examples of quantales: the first six are linear,
  commutative, and affine quantales. The last one, \ie, $\PM(X^2)$, is
  a non-linear, non-commutative, and non-affine quantale (unless $X$
  is trivial).
  The quantale construction $Q/\qI$ returns always an affine quantale
  and preserves the linear and commutative properties, while the
  constructions $\prod_{j \in J}Q_j$ and $\Po(P,Q)$ preserve the
  affine and commutative properties.
\begin{enumerate}
\item $\RQ_+$ is the quantale of non-negative real numbers extended
  with $\infty$ (see~\cite{lawvere1973metric}), with order $x\qle
  y\defiff x\geq y$ and $x\qT y\defeq x+y$.  In particular,
  $\qb=\infty$, $\qI=\qt=0$, and $\qJ S=\inf S$ for every $S\subseteq
  [0, \infty]$.

\item $\RQ_\qm$ is similar to $\RQ_+$, except
  $x\qT y\defeq x\qm y=\max(x,y)$. Thus, $\RQ_\qm$ is a locale and
  $[x,z]=z$ if $x\leq z$ else $0$. The rest, {\ie}, $\qb$,
  $\qI= \qt = 0$, $\qJ S$, and $\qM S$, are the same as in $\RQ_+$.

\item $\NQ_+$ is the sub-quantale of $\RQ_+$ with carrier
  the set of natural numbers extended with $\infty$.  $\NQ_\qm$ is
  the \emph{sub-locale} of $\RQ_\qm$ with the same carrier as
  $\NQ_+$.
  
\item $\BQ$ is the sub-quantale of $\RQ_+$ with carrier
  $\setf{0,\infty}$. $\BQ$ is a locale.

\item $[0,1]_*$ is the probability quantale, \ie, the unit
  interval with the natural order and $x\qT y\defeq x*y$.

\item $[0,1]_\qm$ is fuzzy quantale, which is like $[0,1]_*$, but
  $x\qT y\defeq \min(x,y)$, namely the interpretation of conjunction
  in fuzzy logic.

\item $Q/\qI$ is the sub-quantale of $Q$ with carrier
  $\setb{x \in Q}{x\qle\qI}$, thus $\qI$ is the top element
  of~$Q/\qI$.

\item
  $\prod_{j \in J}Q_j$ is the product of the quantales $Q_j$, with
  $\qle$ and $\qT$ defined pointwise.

\item
  The hom-preorder $\Po(P,Q)$ is a quantale with $\qT$ defined
  pointwise, when $P$ is a preorder and $Q$ a quantale.
  
\item $(\PM(X^2),\subseteq,\qT)$ is the quantale of relations on $X$
  ordered by inclusion, with $\qI=\setb{(x,x)}{x\in X}$ and $R\qT
  S\defeq \setb{(x,z)}{\exists y \in X.(x,y)\in R, (y,z)\in S}$.
\end{enumerate}
Given an ordered monoid $(P,\qT,\qI)$, \ie, a monoid in $\Pos$, its
dual $(P,\qT,\qI)^{op}$ is $(P^{op},\qT,\qI)$, \ie, the order is
reversed, but the monoid structure is unchanged.  The dual of a
quantale $Q$ is never a quantale (unless $Q$ is trivial), because
$\qb=\qb\qT\qt=\qt$ when $\qT$ distributes also over arbitrary meets.
\end{example}
Quantales are ordered monoids with additional properties.  The
following result shows that every ordered monoid
can be \emph{faithfully embedded} into a quantale.
More precisely, every poset $P$ can be \emph{faithfully embedded} into a
complete lattice $\DM(P)$---the \emph{free complete join-semilattice}
over $P$---and any structure of ordered monoid on $P$ induces a
structure of ordered monoid on $\DM(P)$ satisfying the distributivity
laws for quantales.
\begin{proposition}\label{prop-D-construction}
  Given a poset $P$, a set $I$ and an arrow $o:P^I\to P$ in $\Pos$, let
  \hideEM{HIDE. the results below should generalize to $o:P^I\to P$, where
    $I$ is a poset (not necessarily finite), in this way one can
    handle signatures $\Omega$ with operations of arity in $\Pos$.  }
  \begin{itemize}
  \item $\DM(P)$ be the poset of lower (aka downward closed) subsets
    of $P$ ordered by inclusion, we write $\qLowerSet{A}$ for the
    downward closure $\setb{p'}{\exists p\in A.p'\qle p}$ of
    $A\subseteq P$
  \item $\eta:P\to \DM(P)$ be the monotonic map
    $\eta(p)\defeq \qLowerSet{\setf{p}}$, and
    
  \item $\hat{o}:\DM(P)^I\to \DM(P)$ be the monotonic map
    $\hat{o}(\Theta)\defeq \qLowerSet{\setb{o(\theta)}{\theta\in\prod_{i
        \in I}
      \Theta(i)}}$
  \end{itemize}
  then the following properties hold
  \begin{enumerate}
  \item $\DM(P)$ is a complete lattice, where $\qJ
    S\defeq\bigcup S$ and $\qM S\defeq\bigcap S$ for every $S\subseteq
    \DM(P)$
  \item $p_0\qle p_1\iff\eta(p_0)\subseteq\eta(p_1)$ for every $p_0,p_1\in P$
  \item $\hat{o}(\eta\circ\theta)=\eta(o(\theta))$ for every $\theta\in P^I$
  \item
    $\hat{o}(\qJ S(i) \mid i\in
    I)=\qJ\setb{\hat{o}(\Theta)}{\Theta\in\prod_{i \in I}S(i)}$ for
    every $S\in\PM(\DM(P))^I$, \ie, $\hat{o}$ distributes over
    joins\footnote{If $I=\emptyset$, then $S$ is the empty map and
      distributivity becomes
      $\hat{o}(\emptyset)=\qJ\setf{\hat{o}(\emptyset)}$, which is
      trivially true.}
  \item if $\qm$ is the binary meet operation on $P$, then $\hat{\qm}$
    is the binary meet operation on $\DM(P)$.
  \end{enumerate}
\end{proposition}
\begin{remark}
The map $\eta:P\to \DM(P)$ is the universal arrow into the \emph{free
complete join-semilattice} over $P$, and corresponds to the Yoneda
embedding for $\BQ$-enriched categories\footnote{The poset/ $\DM(P)$
is isomorphic to the hom-poset $\Po(P^{op},\BQ)$ of monotonic maps
from $P^{op}$ to $\BQ$, which is actually a locale.}  (see~\cite[Sec
  2.4]{kelly1982basic}), where $\BQ$ is the locale given in
Example~\ref{ex:quantale}.
$\DM$ is part of a $\Pos$-enriched monad on $\Pos$, in
particular the extension $f^*:\DM(P)\to \DM(P')$ of a monotonic map
$f:P\to \DM(P')$ is $f^*(A)\defeq\qJ\setf{f(p) \mid p\in A}$, and the
category of EM-algebras for $\DM$ is (isomorphic to) the category of
complete lattices and join-preserving maps, as there is at most
one $\DM$-algebra structure on a poset $P$, namely the map $\qJ:\DM(P)\to
P$, which is also the left-adjoint of $\eta:P\to \DM(P)$ in $\Pos$.
\end{remark}

In~\cite{Adamek2021} \emph{strongly finitary} monads on $\Pos$ are
characterized in terms of varieties of $\Omega$-algebras in $\Pos$
satisfying a set of $\Omega$-inequations $e_1\leq e_2$, where a
signature $\Omega$ is a set $O$ of operation symbols and a map $\#$
assigning to each $o\in O$ its arity $\#o\in\omega$ (viewed as a
discrete poset), and a $\Omega$-algebra is an arrow $\alpha:K_\Omega
A\to A$ in $\Pos$, where $K_\Omega$ is the polynomial functor
$K_\Omega X=\coprod_{o\in O} X^{\#o}$ on $\Pos$.

The following result identifies a class of inequations preserved by
the construction $\DM(P)$, \ie, it gives sufficient conditions (on the
presentation of) a strongly finitary monad $T$ on $\Pos$ for having a
lifting of $\DM$ along the forgetful functor $U:\Pos^T\to\Pos$, where
$\Pos^T$ is the category of EM-algebras for $T$.
\begin{proposition}\label{prop-D-alg}
  Given a $\Omega$-algebra $\Alg=(P,(o:P^{\#o}\to P \mid o\in O))$ in
  $\Pos$, let $\hat{\Alg}$ be the $\Omega$-algebra
  $(\DM(P),(\hat{o} \mid o\in O))$ in $\Pos$, then $\eta:P\to \DM(P)$ is a
  $\Omega$-algebra homomorphism, moreover if $\Alg$ satisfies the
  $\Omega$-inequation $e_1\leq e_2$, then also $\hat{\Alg}$ does,
  provided that the term $e_1$ is linear (\ie, each variable
  in $e_1$ occurs once) and each variable in $e_2$ occurs in $e_1$.
\end{proposition}
\begin{proof}
If $e$ is a linear term, $X$ is its set of variables and $o:P^X\to P$
is its interpretation in $\Alg$, then its interpretation in $\hat{\Alg}$
is $\hat{o}$.
Every term $e$ is the substitution instance $e'[f]$ of a linear term
$e'$, which is equal to $e$ except for the variables, by applying the
substitution $f:X'\to X$ (necessarily unique) mapping a variable
occurrence in $e'$ to the variable $x$ in the corresponding occurrence
of $e$.  Therefore the interpretation of $e$ in $\Alg$ is $P^f:P^X\to
P^{X'}$ followed by the interpretation $o':P^{X'}\to P$ of $e'$.

If $X_i$ is the set of variables in $e_i$, then (by
the assumption on $e_1\leq e_2$) $e_1\equiv e'_1$, $X_2\subseteq X_1$
and $e_2\equiv e'_2[f]$ with $e'_2$ linear term and $f:X'_2\to X_2$
substitution.
Let $o_i:P^{X'_i}\to A$ be the interpretation of $e'_1$ in $\Alg$,
then
\begin{enumerate}
\item the interpretation of $e_1$ in $\Alg$ is $o_1:P^{X_1}\to P$,
  because $e_1\equiv e'_1$
\item the interpretation of $e_2$ is $\theta_2\mapsto
  o_2(\theta_2\circ f)$ in $P^{X_2}\to P$, because $e_2\equiv
  e_2'[f]$
\item\label{alg:fact:3} $o_1(\theta_1)\leq o_2(\theta_1\circ f)$ for every $\theta_1\in
  P^{X_1}$, because $\Alg$ satisfies $e_1\leq e_2$ (and $X_2\subseteq
  X_1$)
\item the interpretation of $e_1$ in $\hat{\Alg}$ is
  $\hat{o}_1:\DM(P)^{X_1}\to \DM(P)$, because $e_1$ is linear
\item the interpretation of $e_2$ in $\hat{\Alg}$ is $\hat{o}_2\circ
  \DM(P)^f:\DM(P)^{X_2}\to \DM(P)$, because $e_2\equiv e'_2[f]$ with $e'_2$
  linear.
\end{enumerate}

We now prove that $\hat{\Alg}$ satisfies $e_1\leq e_2$, \ie,
$\hat{o}_1(\Theta)\subseteq\hat{o}_2(\Theta\circ f)$
for every $\Theta\in \DM(P)^{X_1}$:

\begin{eqnarray*}
  & & \hat{o}_1(\Theta)\\
  (\text{by distributivity of $\hat{o}_1$})  & = &
                                                    \qLowerSet{\setf{o_1(\theta_1)
                                                    \mid
                                                    \theta_1\in\prod_{x
                                                    \in X_1} \Theta(x)}}
  \\
  (\text{by item~\ref{alg:fact:3} above}) & \subseteq  & \qLowerSet{\setf{o_2(\theta_1\circ f) \mid \theta_1\in\prod_{x \in X_1}
                                                         \Theta(x)}} \\
(\text{because
                                            $\theta_1\circ f\in \prod_{x \in X'_2} \Theta(f(x))$ when
                                            $\theta_1\in\prod_{x \in
  X_1} \Theta(x)$})  & \subseteq & \qLowerSet{\setf{o_2(\theta'_2) \mid \theta'_2\in\prod_{x \in X'_2}
                                   \Theta(f(x))}} \\
(\text{by distributivity of $\hat{o}_2$}) & = & \hat{o}_2(\Theta\circ f).  
\end{eqnarray*}
\end{proof}

Examples of (in)equations preserved by the $\hat{\Alg}$ construction
are: $x\qT\qI=x$, $x=\qI\qT x$, $x\qT y=y\qT x$, $x\qT (y\qT z)=(x\qT
y)\qT z$, $x\qT y=x\qm y$, $x\leq\qI$ and $x\leq x\qT x$.
Inequations that may not be preserved are: $\qI\leq x$, $x\qT x\leq
x$ and $x\qT x\leq \qI$.

\begin{example}\label{ex:cost-quantale}
  We consider quantales arising in the analysis of algorithms, where
  algorithms are identified with multi-tape deterministic Turing
  Machines (TM) accepting/rejecting strings written in a finite
  alphabet $A$.
  \begin{itemize}
    \item The size $s(w)$ of an input string $w$ is a value in the
      quantale $\NQ_+$, namely the length of the string $w$.
      The size of an infinite string is $\infty$, and the size of the
      concatenation of two strings is the sum of their sizes.

    \item The time (\ie, the number of steps) taken by a TM on a
      specific input $w$ is again a value in $\NQ_+$.
      A TM failing to terminate on $w$ takes time $\infty$, and the
      time taken for executing two TMs sequentially on $w$ is the sum
      of the times taken by each TM (plus a linear overhead for
      copying $w$ on two separate tapes, so that the two TMs work on
      disjoint sets of tapes).
    \end{itemize}
    The time complexity associated to a TM typically depends on the
    input (or the input size).  Such a cost should be drawn from a
    quantale reflecting this dependency, namely, a \emph{higher-order}
    quantale\footnote{This resembles higher-order metrics used to
    compare functional programs~\cite{DalLagoGY19,Pistone21}.}.  We now
    describe some quantales of this kind from the most concrete to the
    most abstract.
  \begin{enumerate} 
  \item The most concrete quantale is ($\Po(A^*,\NQ_+)$ or
    equivalently) $\NQ_+^{A^*}$, \ie, the product of $A^*$ copies of
    $\NQ_+$. An element $t\in\NQ_+^{A^*}$ maps each finite input $w\in
    A^*$ to the time taken by the TM on $w$.
    
  \item A first abstraction involves replacing $t\in\NQ_+^{A^*}$ with
    $T\in\NQ_+^\omega$, where $T(n)$ is the best upper-bound for the
    time taken by the TM on inputs of size $n$, \ie,
    $T(n)=\max\setb{t(w)}{s(w)=n}$.
    
  \item By the linear speed-up theorem, time complexity is usually
    expressed using $O$-notation, \ie, $T\in\NQ_+^\omega$ is
    replaced with the subset $O(T)$ of $\NQ_+^\omega$ such that $T'\in
    O(T)\iff \exists n_0, C \in \omega. \forall n\geq n_0.T'(n)\leq
    C*T(n)$.
   
    If we replace $\NQ_+^\omega$ with the partial order $L_O$ of
    $O$-classes $O(T)$ ordered by reverse inclusion, we get a
    distributive lattice (\ie, finite meets distribute over finite
    joins\footnote{In this case, the dual property also holds, {\ie},
    finite joins distribute over finite meets.}), the top element is
    $O(\lambda n.0)$, the bottom element is $O(\lambda n.\infty)$, the
    join and the meet are
    $$\begin{array}{l}
     O(T_1)\qj O(T_2)=O(T_1)\cap O(T_2)=O(T_1\qj T_2)=O(\min(T_1,T_2))\\
      O(T_1)\qm O(T_2)=O(T_1\qm T_2) = O(\max(T_1,T_2)) = O(T_1+T_2).    
    \end{array}$$
    The lattice $L_O$ is distributive, because the complete lattice
    $\NQ_+^\omega$ is,  but $L_O$ is not complete.
    For instance, let $T_k(n) \defeq n^k$. Then, the set
    $S \defeq \setb{O(T_k)}{k \in \omega}$ does not have a meet. In
    fact, given $T \in \NQ_+^\omega$ such that
    $\forall k \in \omega.\ T_k \in O(T)$, let
    $\hat{T}(n) \defeq \lceil T(n) / (n+1) \rceil$, where
    $\lceil \cdot \rceil$ is integer round up, then we have
    $\forall k \in \omega.\ T_k \in O(\hat{T})$, $\hat{T} \in O(T)$,
    but $T \notin O(\hat{T})$.
    However, there is a construction which turns a
    distributive lattice $L$ into the \emph{free locale} $I(L)$ over
    $L$ (see~\cite[page 69]{Johnstone86}), \ie,
    the poset of \emph{ideals} in $L$ ordered by inclusion,
    and the embedding $x\mapsto\qLowerSet{x}$ from $L$ to $I(L)$
    preserves finite meets and finite joins.

\item A simpler way to obtain a locale is to take the subset of $L_O$
  consisting of the $O(n^k)$ with $k\in\omega+1$, where
  $O(n^\omega)=O(\lambda n.\infty)$.  This locale is isomorphic to
  $\NQ_\qm$, namely $k\in\NQ_\qm$ corresponds to $O(n^k)$.
  \end{enumerate}
\end{example}

\subsection{Continuous Quantales}

A quantale is \emph{continuous} when its underlying complete lattice
is continuous.  By restricting to continuous quantales we can recast
the \emph{epsilon-delta} characterizations of continuous and uniformly
continuous maps in the context of quantale-valued metric spaces, and
relate such metric spaces to topological spaces.

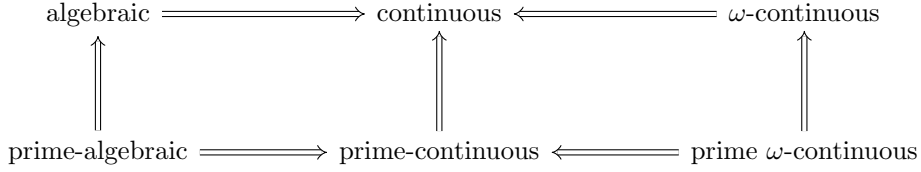
\begin{figure}[tb]
\begin{equation*}
  \begin{tikzcd}[row sep = huge, column sep = huge]
    \text{algebraic} \arrow[r, Rightarrow] & \text{continuous} &
    \text{$\omega$-continuous} \arrow[l, Rightarrow]\\
    \text{prime-algebraic} \arrow[u, Rightarrow] \arrow[r, Rightarrow]
    & \text{prime-continuous} \arrow[u, Rightarrow] & \text{prime
      $\omega$-continuous} \arrow[u, Rightarrow] \arrow[l, Rightarrow]
  \end{tikzcd} 
\end{equation*}
  \caption{Relations among different types of complete lattices.}
  \label{fig:CL-variants}
\end{figure}

We recall the definitions of continuous lattice and related notions
(see Definition~\ref{def:CL}), which are summarized in
Fig.~\ref{fig:CL-variants}, we state some facts about complete
lattices with these additional properties
(see~\cite{AbramskyJung94-DT,Gierz-ContinuousLattices-2003,Goubault-Larrecq:Non_Hausdorff_topology:2013}),
and prove some properties of quantales, when the underlying complete
lattice has any of these properties.

\begin{definition}
\label{def:way_below}
Given a complete lattice $(Q,\qle)$ and $x,y \in Q$, we say that:
\begin{enumerate}
\item $D \subseteq Q$ is directed $\defiff$ $D$ non-empty and
  $(\forall x,y \in D. \exists z \in D. x,y \qle z)$.
\item $x$ is \textbf{way-below} $y$ (notation $x\ll y$) $\defiff$ for
  any directed subset $D$ of $Q$, $y\qle\qJ D\implies\exists d \in
  D.x\qle d$.  We write $\qwb y$ for the set $\setb{x \in Q}{x\ll y}$.
\item $x$ is \textbf{compact} $\defiff x \ll x$.
 $\KE(Q)$ denotes the set of compact elements in $Q$.
\item $x$ is \textbf{totally-below} $y$ (notation $x\lll y$) $\defiff$
  for any subset $D$ of $Q$, $y\qle\qJ D\implies\exists d \in D.x\qle
  d$.
\item $x$ is completely \textbf{join-prime} $\defiff x \lll x$. 
  $\PE(Q)$ denotes the set of completely join-prime elements in $Q$.
\end{enumerate}
\end{definition}
The following are some basic properties of the way-below and
totally-below relations.
\begin{proposition}\label{thm:wb:prop}
  In any complete lattice $(Q,\qle)$, and for all $x, x_0, x_1 \in Q$:
\begin{enumerate}
\item $x_0\lll x_1\implies x_0\ll x_1\implies x_0\qle x_1$.
  Therefore $\PE(Q)\subseteq\KE(Q)$.
\item $x_0'\qle x_0\ll x_1\qle x_1'\implies x_0'\ll x_1'$.
\item $x_0'\qle x_0\lll x_1\qle x_1'\implies x_0'\lll x_1'$.
\item $\bot\ll x$ and $\qwb x$ is a directed subset of $Q$, more
  precisely $x_0,x_1\ll x\implies x_0\qj x_1\ll x$.
\item $\KE(Q)$ is a directed subset of $Q$.
\end{enumerate}
\end{proposition}

\begin{definition}\label{def:CL}
Given a complete lattice $Q$, we say that:
\begin{enumerate}
\item $Q$ is \textbf{continuous} $\defiff\forall
  y \in Q.y=\qJ\qwb y$.
\item $B\subseteq Q$ is a \textbf{basis} for $Q$ $\defiff\forall y \in Q.
  B\cap\qwb y$ is directed and
  $y=\qJ(B\cap\qwb y)$.
\item $Q$ is \textbf{$\omega$-continuous} $\defiff$ $Q$ has a countable basis.
\item $Q$ is \textbf{algebraic} $\defiff\forall y\in Q.y =
  \qJ\setb{x\in\KE(Q)}{x\ll y}$, or equivalently $\KE(Q)$ is a basis.
\item $Q$ is \textbf{prime-continuous} $\defiff\forall
  y \in Q.y=\qJ\setb{x \in Q}{x\lll y}$.
\item $B\subseteq Q$ is a \textbf{prime basis} for $Q$ $\defiff\forall y \in Q.
  y=\qJ\setb{x \in B}{x\lll y}$.
\item $Q$ is \textbf{prime $\omega$-continuous} $\defiff$ $Q$ has a
  countable prime basis.
\item $Q$ is \textbf{prime-algebraic} $\defiff\forall y \in
  Q.y=\qJ\setb{x \in \PE(Q)}{x\lll y}$, or equivalently $\PE(Q)$ is a
  prime basis.
\end{enumerate}
$\CL$ denotes the cartesian closed category of continuous lattices and
Scott continuous maps.
\end{definition}
A complete lattice $Q$ is continuous exactly when it has a basis, every
basis includes $\KE(Q)$, but $\KE(Q)$ is a basis only when $Q$ is algebraic.
Similarly, a complete lattice $Q$ is prime-continuous exactly when it
has a prime basis, every prime basis includes $\PE(Q)$, but $\PE(Q)$
is a prime basis only when $Q$ is prime-algebraic.

Every prime-continuous lattice is clearly continuous.  From the
properties stated in Proposition~\ref{thm:wb:prop} one can easily
derive the following relations between prime basis and basis in prime-continuous lattices.
\begin{corollary}
  \label{cor:basis_prime_basis}
  Given a prime-continuous lattice $Q$ (and therefore also continuous)
  \begin{enumerate}
  \item If $B$ is a basis for $Q$, then $B$ is also a prime basis for
    $Q$.
\item If $B$ is a prime basis for $Q$, then the closure $B_\qj$ of $B$
  under finite joins is a basis for $Q$.
\end{enumerate}
\end{corollary}

Prime-continuous lattices are a special case of continuous lattices
with a stronger notion of approximation given by the totally-below
relation $\lll$, compared with that given by the way-below relation in
continuous lattices. The framework of the current article is based on
continuous lattices, and the main reason for discussing
prime-continuous lattices is the extensive metrization result of
Section~\ref{subsec:Extensive_Metrization}. While not every topology
is induced by a classical metric,
Kopperman~\cite{Kopperman:All_topologies_metric:1988} showed that
every topology is induced by a \emph{generalized} metric.  We will also
show that every topology is induced by a generalized metric based on
continuous quantales and open balls defined via the way-below
relation. For that, we will rely on the extensive metrization result
in~\cite[Example~11]{Cook_Weiss:quantales_Lip:2022}, which is based on
prime-algebraic locales and open balls defined via the
totally-below relation.

Prime-continuous lattices can be characterized in various ways, including:
\begin{proposition}[{see~\cite[Thm 7.13]{AbramskyJung94-DT},
     ~\cite[Exercise~8.3.47]{Goubault-Larrecq:Non_Hausdorff_topology:2013}
      and~\cite[Exercise 7.3.19(1)]{AbramskyJung94-DT}}]\ \\
  If $Q$ is a complete lattice, then the following assertions are equivalent
  \begin{enumerate}
  \item $Q$ is prime-continuous
  \item $Q$ and $Q^{op}$ are continuous and distributive (\ie, finite
    joins distribute over finite meets)
  \item $Q$ is completely distributive (\ie, arbitrary joins
    distribute over arbitrary meets).
  \end{enumerate}
\end{proposition}
From the above characterizations we can derive the following properties
of prime-continuous lattices.
\begin{corollary}\label{cor:PCL:local}
  If $Q$ is a prime-continuous lattice, then $Q$ is a locale and
  $Q^{op}$ is prime-continuous.
\end{corollary}
prime-continuous lattices (like complete lattices) are
\emph{self-dual}, \ie, they are closed under taking the dual $P^{op}$
(in $\Po$), while all other properties of complete lattices introduced
in Definition~\ref{def:CL} are not self-dual.
\begin{corollary}
  Every linear complete lattice is prime-continuous.
\end{corollary}
\begin{proof}
  If $Q$ is linear complete lattice, then it is continuous (see
 ~\cite[Example~I-1.7, page 55]{Gierz-ContinuousLattices-2003}, where
  linear complete lattices are called complete chains) and
  distributive.  But also $Q^{op}$ is a linear complete lattice, and
  thus continuous and distributive.  therefore, we can conclude that
  $Q$ is prime-continuous.
\end{proof}

Complete lattices are closed under small products computed in $\Po$,
and if $Q$ is a complete lattice, then for any preorder $P$ the hom-preorder
$\Po(P,Q)$ is a complete lattice.
Also complete lattices with any of the properties in
Definition~\ref{def:CL} enjoy similar closure properties.
\begin{proposition}\label{prop:cont_quantales}
  Continuous, prime-continuous, algebraic and prime-algebraic lattices
  are closed under small products (in $\Po$), while
  $\omega$-continuous and prime $\omega$-continuous lattices are
  closed under countable products.
\end{proposition}
\begin{proof}
  The claims follow from the following facts:
\begin{itemize}
\item The projection $\pi_i:(\prod_{j \in J}Q_j)\to Q_i$ preserves
  joins and meets, because they are computed pointwise, therefore $\pi_i$
  has left- and right-adjoints $\bot_i,\top_i:Q_i\to(\prod_{j \in
    J}Q_j)$ given by $\bot_i(q)_j\defeq\mbox{$q$ if $i=j$ else
    $\bot_j$}$ and $\top_i(q)_j\defeq\mbox{$q$ if $i=j$ else
    $\top_j$}$.
\item If $\forall j \in J.B_j$ is a basis for $Q_j$, then
  $B\defeq\setb{\bot_i(q)}{i\in J\land q\in Q_i}$ is a sub-basis for
  $\prod_{j \in J}Q_j$, \ie, the set of finite joins of elements in $B$
  form a basis.
\item If $\forall j \in J.B_j$ is a prime basis for $Q_j$, then
  $B\defeq\setb{\bot_i(q)}{i\in J\land q\in Q_i}$ is a prime basis for
  $\prod_{j \in J}Q_j$.
\end{itemize}
\end{proof}

\begin{proposition}\label{prop:cont_quantales_2}
  If the complete lattice $Q$ is continuous, prime-continuous,
  algebraic or prime-algebraic, then so is $\Po(P,Q)$ for any preorder
  $P$.
  If the complete lattice $Q$ is $\omega$-continuous or prime
  $\omega$-continuous, then so is $\Po(P,Q)$ for any countable
  preorder $P$.
\end{proposition}
\begin{proof}
  The claims follow from the following facts:
\begin{itemize}
\item The evaluation map $\pi_p:\Po(P,Q)\to Q$ such that
  $\pi_p(f)=f(p)$ preserves joins and meets, and its left-
  and right-adjoints $\bot_p,\top_p:Q\to\Po(P,Q)$ are
  $\bot_p(q)(p')\defeq\mbox{$q$ if $p\qle p'$ else $\bot$}$ and
  $\top_p(q)(p')\defeq\mbox{$q$ if $p'\qle p$ else $\top$}$.
\item If $B$ is a basis for $Q$, then $\setb{\bot_p(q)}{p\in P\land
  q\in B}$ is a sub-basis for $\Po(P,Q)$.
\item If $B$ is a prime basis for $Q$, then $\setb{\bot_p(q)}{p\in
  P\land q\in B}$ is a prime basis for $\Po(P,Q)$.
\end{itemize}
\end{proof}
\begin{remark}
  In general, the map $\bot_p(q)$ is not in the hom-set $\CL(P,Q)$,
  unless $p$ is a compact element of the continuous lattice
  $P$. Furthermore, the inclusion $\CL(P,Q)\hookrightarrow\Po(P,Q)$ preserves
  joins, but not meets.
\end{remark}

We recall some well-known facts about Scott topology on complete
lattices.
\begin{itemize}
\item A subset $O$ of a complete lattice $Q$ is \emph{Scott open} if
  it has the following properties:
  \begin{enumerate}
  \item It is an upper set, \ie, if $x \in O$ and $x \sqsubseteq y$,
    then $y \in O$.
  \item It is inaccessible by directed joins, \ie, 
    if $D \subseteq Q$ is directed and $\qJ D \in O$, then
    $D \cap O \neq \emptyset$.
  \end{enumerate}
\item
  A map $f: Q \to Q'$ between complete lattices is Scott
  continuous if it is continuous with respect to the Scott topologies
  on $Q$ and $Q'$.
  Scott continuity can be recast in purely order-theoretic terms,
  namely $f: Q \to Q'$ is Scott continuous if and only if it preserves
  directed joins, \ie, $f(\qJ D) = \qJ f(D)$ for every directed $D
  \subseteq Q$ (see~\cite[Proposition~4.3.5]{Goubault-Larrecq:Non_Hausdorff_topology:2013}).
  \item
  If $Q$ is a continuous lattice, then for any $x\in Q$ the set $\qwa
  x \defeq \setb{y \in Q}{x \ll y}$ is Scott open, and for any basis
  $B$ for $Q$ the set $\setb{\qwa b}{b \in B}$ is a base for the Scott
  topology on $Q$ (see~\cite[Proposition~2.3.6]{AbramskyJung94-DT}).
\end{itemize}

Continuous and prime-continuous lattices have the following
interpolation property.
\begin{lemma}[{see~\cite[Lemma~2.2.15]{AbramskyJung94-DT}}]
  \label{lemma:inter:cl}\ 
  \begin{itemize}
  \item If $Q$ is a continuous lattice, then $q_1 \ll
    q_2\implies\exists q \in Q.q_1\ll q\ll q_2$.
  \item If $Q$ is a prime-continuous lattice, then $q_1 \lll
    q_2\implies\exists q \in Q.q_1\lll q\lll q_2$.
  \end{itemize}
\end{lemma}
In continuous quantales also the following interpolation properties
hold.  In prime-continuous quantale also the properties where $\ll$ is
replaced by $\lll$ hold (and the proofs are similar).
\begin{lemma}
  \label{lemma:inter:cq}
  If $Q$ is a continuous quantale, then
  \begin{enumerate}
  \item $q_1 \ll q_2\implies\exists q\ll\qI.\ q_1\ll q_2\qT q$.
  \item $q_1 \ll q_2\implies\exists q\ll\qI.\ q_1\ll q\qT q_2$.
  \end{enumerate}
\end{lemma}
\begin{proof}
Note that:
\begin{eqnarray*}
  q_2 & = &q_2 \qT \qI \\  
  (\text{by continuity of $Q$})& = & q_2 \qT (\qJ\setb{q \in Q}{q\ll \qI}) \\
  (\text{by distributivity for $\qT$})& = & \qJ \setb{q_2 \qT q}{q\ll \qI}.
\end{eqnarray*}
\begin{itemize}
\item $\setb{q_2 \qT q}{q\ll \qI}$ is directed, because $\qwb\qI$ is
  directed and $\qT$ is monotone.
\item $q_1\ll q' \ll q_2$ for some $q'$, because of the interpolation
  property for $\ll$.
\item $q'\qle q_2 \qT q$ for some $q \ll \qI$, because $q'\ll
  q_2=\setb{q_2 \qT q}{q\ll \qI}$, thus
\item $q_1\ll q_2 \qT q$, by Proposition~\ref{thm:wb:prop}.
\end{itemize}
The second implication can be proved in a similar way.
\end{proof}

\begin{lemma}\label{lemma:inter:cq2}
If $Q$ is a continuous quantale, the following hold: 
\begin{enumerate}
\item If $q \ll q_1\qT q_2$ then there is $q' \ll q_2$ such that $q \ll q_1 \qT q'$. 
\item If $q \ll q_1\qT q_2$ then there is $q' \ll q_1$ such that $q \ll q' \qT q_2$. 
\end{enumerate}
\end{lemma}
\begin{proof}
  It follows immediately from the following facts:
  the interpolation property for $\ll$ (see Lemma~\ref{lemma:inter:cl}),
  $q_1$ and $q_2$ are the (directed) join of the elements way-below them, 
  $\qT$ preserves arbitrary join.
\end{proof}

\begin{example}\label{ex:continuous-quantale}
The quantales in Example~\ref{ex:quantale} have the following properties:
\begin{itemize}
\item $\NQ_+$, $\NQ_\qm$, and $\BQ$ are linear and (prime)
  $\omega$-algebraic.  The way-below relation is $x\ll y\iff
  x\qle y$, all elements in these quantales are compact (and
  completely join-prime except the least element).
  
\item $\RQ_+$ and $\RQ_\qm$ are linear and (prime)
  $\omega$-continuous, \eg, the set of positive rational numbers is a
  prime basis.  The totally-below relation is $x\lll y\iff x\qlt y$,
  and the only compact element is $\bot = \infty$.

  Similarly for $[0,1]_*$ and $[0,1]_\qm$, \eg, the set of rational
  numbers in $(0,1)$ is a prime basis.

\item $\PM(X^2)$ is prime-algebraic, and its completely join-prime
  elements are the singleton, since the totally-below relation in a
  complete lattice $\PM(Y)$ ordered by inclusion is $A\lll B\iff
  A\subseteq B$ and $A$ is a singleton.
\end{itemize}
The quantales in Example~\ref{ex:cost-quantale}  have the following properties:
\begin{itemize}
\item $\NQ_+^{A^*}$ and $\NQ_+^\omega$ are prime $\omega$-algebraic, because
  they are products of countably many copies of $\NQ_+$.
\item The locale $I(L_O)$ of ideals in the distributive lattice $L_O$
  is algebraic, since its compact elements are the principal ideals in
  $L_O$ (which are in bijective correspondence with the elements of
  $L_O$), but it is not prime-continuous, which we prove by
  contradiction.

  If $I(L_O)$ were prime-continuous, then by
  Corollary~\ref{cor:basis_prime_basis}, the base $\KE(I(L_O))$ would
  also be a prime basis for $I(L_O)$.  We prove that
  $\qLowerSet{O(T)} \lll \qt \iff \lambda n. \infty \in O(T)$ for any
  $O(T) \in L_O$.
  In other words, the only element totally-below the top element
  $\qt=\qLowerSet{O(\lambda n.0)}$ is the bottom element
  $\qb=\qLowerSet{O(\lambda n.\infty)}$.
  Since $\qLowerSet:L_O\to I(L_O)$ preserves finite joins and finite meets,
  and $\KE(I(L_O))$ is the image of $L_O$ along $\qLowerSet$, we may work in
  $L_O$.
  If $O(T) \neq O(\lambda n.\infty)$, then there exists a strictly
  increasing sequence of natural numbers $(n_k \mid k\in\omega)$ such that
  $\forall k \in \omega. T(n_k) \neq \infty$.
  If $T_1, T_2\in \NQ_+^\omega$ are given by:
  \begin{equation*}
    T_1(n) \defeq
    \begin{cases}
      \infty , & \text{if } \exists k \in \omega.\ n_{2k} = n,\\
      0, & \text{otherwise},\\
    \end{cases}
    \qquad
    T_2(n) \defeq
    \begin{cases}
      \infty , & \text{if } \exists k \in \omega.\ n_{2k+1} = n,\\
      0, & \text{otherwise},\\
    \end{cases}
  \end{equation*}
then $O(T_1)\qj O(T_2)=O(\min(T_1,T_2))=O(\lambda n.0)$ and $T_1,T_2
\notin O(T)$, \ie, $O(T) \not \qle O(T_1), O(T_2)$, which contradicts
the assumption $\qLowerSet{O(T)} \lll \qt$.
\end{itemize}
The complete lattice $\DM(P)$ in Proposition~\ref{prop-D-construction}
is prime-algebraic, thus it is a locale by
Corollary~\ref{cor:PCL:local}.  Its completely join-prime elements are
the $\qLowerSet{a}$ with $a\in P$.
\end{example}

\begin{example}\label{ex:conv-qnt}
  Given an ordered commutative monoid $(P,\qT_P,\qI_P)$ and a
  commutative quantale $(Q,\qT_Q,\qI_Q)$, we get a commutative
  quantale on the complete lattice $\Po(P,Q)$ by taking Day
  convolution product as the tensor:
  \begin{equation*}
  (f_1\conv f_2)(x) \defeq \qJ_{x_1\qT_P x_2 \qle x} f_1(x_1)\qT_Q f_2(x_2)  
  \end{equation*}
  with the unit given by $\epsilon(x)\defeq\mbox{$\qI_Q$ if
    $\qI_P\qle x$ else $\qb_Q$}$.
The fact that we get a commutative quantale follows from general
properties of Day convolution product established in the setting of
enriched categories (see~\cite{kelly1982basic}).

When $P$ and $Q$ are prime-continuous lattices,\footnote{We are only
  assuming $(P,\qT_P,\qI_P)$ to be an ordered monoid, not a quantale,
  therefore the distributivity laws may fail.} we may consider the
prime-continuous lattice $\PCL(P,Q)$ of join-preserving maps from $P$
to $Q$.
In general, $\PCL(P,Q)$ as a subset of $\Po(P,Q)$ is closed under
arbitrary joins computed in $\Po(P,Q)$, but it may not be closed under
$\conv$, and $\epsilon$ may not be in $\PCL(P,Q)$.  In other words,
$\PCL(P,Q)$ may fail to be a sub-quantale of
$(\Po(P,Q),\conv,\epsilon)$.

Day convolution, however, restricts to $\PCL(P,Q)$ under the following
condition on $P$:
\begin{quote}
\Star for all $x_1,x_2,y_1,y_2 \in P$, if $x_1\lll y_1$ and $x_2\lll
y_2$, then $x_1\qT_P x_2\lll y_1\qT_P y_2$.
\end{quote}
Indeed, if $f,g : \PCL(P,Q)$, $A\subseteq P$ and $x = \qJ A$, then
$\qJ_{z\in A} (f\conv g)(z) \qle (f\conv g)(x)$, because $f\conv g$
is monotone.
While for the other inequality, we have
\begin{equation*}
(f\conv g)(x) 
  = \qJ_{x_1\qT_P x_2\qle x} f(x_1) \qT_Q g(x_2)
  = \qJ_{x_1\qT_P x_2\qle x} (\qJ_{y_1\lll x_1}\ \qJ_{y_2\lll x_2} f(y_1)\qT_Q g(y_2)
).  
\end{equation*}
Therefore, given $x_1\qT_P x_2\qle x$, $y_1\lll x_1$ and
$y_2\lll x_2$, from \Star we obtain
$y_1\qT_P y_2 \lll x_1\qT_P x_2 \qle x$.
Hence, there exists $y \in A$ such that $y_1\qT_P y_2 \qle y$, which
implies $f(y_1)\qT_Q g(y_2) \qle (f\conv g)(y) \qle \qJ_{z\in A}
(f\conv g)(z)$.

In general, \Star does not imply that the unit is in $\PCL(P,Q)$, \eg:
if $P=\RQ_+$, then $\epsilon(x)=\mbox{$\qI_Q$ if $x=0$ else $\qb_Q$}$
is not even Scott continuous (unless $Q$ is trivial); if $P$ is
$\RQ_+^{op}$ (which is not a quantale), then $\epsilon=\lambda
x:P.\qI_Q$ is not join-preserving (unless $Q$ is trivial).

To ensure that $\conv$ restricted to $\PCL(P,Q)$ has a unit in
$\PCL(P,Q)$, a further condition on $P$ is needed:
\begin{quote}
\SStar for all $x,y \in P$, if $x\lll y$, then there are
$x_1,x_2\ggg \qI_P$ such that $x_1\qT_P x \lll y$ and $x\qT_P x_2 \lll y$.
\end{quote}
In this case we can prove that the unit is given by the join-preserving map
$\epsilon^\lll(x)\defeq\mbox{$\qI_Q$ if $\qI_P\lll x$ else $\qb_Q$}$.
First, $(\epsilon^\lll\conv f) \qle f$ holds, because $x_1\qT_P x_2
\qle x$ implies $\epsilon^\lll(x_1)\qT_Q f(x_2)=\mbox{$f(x_2)$ if
  $\qI_P\lll x_1$ else $\qb_Q$}\qle f(x)$.
For the other inequality, since $f$ is join-preserving, it suffices to
show that $y \lll x$ implies $f(y) \qle (\epsilon^\ll\conv f)(x)$.
By \SStar, we get $x'\qT_P y \lll x$ for some $x'\ggg\qI_P$, hence
$f(y) = \epsilon^\lll(x')\qT_Q f(y) \qle (\epsilon^\lll\conv f)(x)$.

A key instance of this construction is obtained when $P=\RQ_+^{op}$
and $Q=[0,1]_*$ (see Example~\ref{ex:quantale}).  In this case the
prime-continuous quantale $(\PCL(P,Q),\conv,\epsilon^\lll)$ is the
distance distributions quantale for probabilistic metric spaces (see
\cite{FlaggK97,Hofmann_Reis:Prob_Met_Spaces:2013,HofmannST14,Cook_Weiss:quantales_Lip:2022}).
Indeed, the maps in $\PCL(P,Q)$ are exactly the functions
$f:[0,\infty]\to [0,1]$ such that $f(x) = \sup_{y < x} f(y)$, or
equivalently $f(x)=P(X<x)$ for a random variable $X$ on $[0,\infty]$.

More generally, when $P$ and $Q$ are continuous lattices, one can
define \emph{mutatis mutandis} a continuous quantale
$(\CL(P,Q),\conv,\epsilon^\ll)$, provided $P$ satisfies the variants
of \Star and \SStar with the totally-below relation $\lll$ replaced by
the way-below relation $\ll$.
\end{example}

\subsection{Quantale Morphisms}

There are several notions of quantale morphism, depending on the
structure considered relevant:
\begin{itemize}
  \item if the relevant structure is the complete lattice, then
    appropriate morphisms are monotonic maps, Scott continuous maps
    (\ie, preserving directed joins), or join-preserving maps (\ie,
    preserving all joins);
  \item if the relevant structure is the ordered monoid, then
    appropriate morphisms are analogues of lax- or strict-monoidal
    functors.
\end{itemize}
\begin{definition}\label{def:qmaps}
  A monotonic map $h:Q\to Q'$ between quantales is called:
  \begin{itemize}
  \item \textbf{lax-unital} $\defiff$ $\qI'\qle'h(\qI)$;
  \item \textbf{lax-monoidal} $\defiff$ $\qI'\qle'h(\qI)$ and
    $\forall x,y \in Q.h(x)\qT'h(y)\qle' h(x\qT y)$;
  \item \textbf{strict-monoidal} $\defiff$
        $\qI'=h(\qI)$ and $\forall x,y \in Q.h(x)\qT'h(y)=h(x\qT y)$.
  \end{itemize}
  We write $\Quan$ for the $\Pos$-enriched category of continuous
  quantales and Scott continuous lax-unital maps.  We may restrict to
  sub-categories of $\Quan$, the most relevant ones are given in
  Fig.~\ref{fig:quantale-cats}.
\end{definition}
\begin{figure}[tb]
  \begin{center}
    \begin{tabular}{|r|l|}
      \hline
    \textbf{Restriction on arrows} & \textbf{Sub-category}
    \\\hline \hline
    lax-monoidal&$\Quan_{lm}$
    \\\hline
    strict-monoidal&$\Quan_{sm}$
    \\\hline
    join-preserving&$\Quan_J$
    \\\hline
    identity on $Q$&$Q$
    \\\hline
  \end{tabular}
  \end{center}
  \caption{Sub-categories of $\Quan$.}
  \label{fig:quantale-cats}
\end{figure}

The three properties above make sense for any monotonic map $h:P\to
P'$ between ordered monoids, and the construction in
Proposition~\ref{prop-D-construction} turns $h$ into a join-preserving
map $\DM(h):\DM(P)\to \DM(P')$ between algebraic quantales inheriting
from $h$ any of these properties.

\begin{example}\label{ex:monoidal-maps}
  We give examples of monotonic maps between quantales.
  In the following diagrams%
\begin{tikzcd}
  {} \arrow[r, dashed] &  {}
\end{tikzcd}
means lax-monoidal map,%
\begin{tikzcd}
  {} \arrow[r] &  {}
\end{tikzcd}%
means strict-monoidal map, and $f\dashv g$ means
``$f$ is left-adjoint to $g$''.
\begin{equation*}
  \begin{tikzcd}[row sep = large, column sep = large]
    1 \arrow[r, dashed, "\qt_Q", "\top"', yshift = 1.1ex] & Q
    \arrow[l, "!_Q", yshift = -1.1ex] \arrow[r, dashed, "g", "\top"', yshift =
    1.1ex] & Q/\qI \arrow[l, "f", yshift = -1.1ex] \arrow[r, "g'", "\top"',
    yshift = 1.1ex] & \BQ \arrow[l, "f'", yshift = -1.1ex] 
  \end{tikzcd} \quad
\end{equation*}
where $1$ is a trivial quantale (\ie, a singleton $\{*\}$),
$!_Q$ is the unique map from $Q$ to $1$, 
\begin{itemize}
\item $\qt_Q$ maps $*$ (\ie, the unique element in $1$) to $\qt$;
\item $f$ is the inclusion of $Q/\qI$ into $Q$, and $g$ maps $x$ to
  $x\qm\qI$;
\item $f'$ maps $\qb$ to $\qb$ and $\qt$ to $\qt$, and
  $g'$ maps $\qt$ to $\qt$ and $x\qlt\qt$ to $\qb$;
\end{itemize}
In general, the maps $g$ and $g'$ are not Scott continuous,
but $g$ is Scott continuous when $Q$ is continuous.
Left-adjoints are always join-preserving maps.
\begin{equation*}
  \begin{tikzcd}[row sep = large, column sep = large]
    \NQ_+ \arrow[r, "i"', yshift = -1.1ex] & \RQ_+
    \arrow[l, dashed, "c"', "\top", yshift = 1.1ex]  & \RQ_\qm \arrow[l, dashed, "\id{}"']
    \end{tikzcd}
  \end{equation*}
where $i$ is the inclusion, $c(x)=\lceil x\rceil$ is integer round up
(which is not Scott continuous), and $\id{}$ is the identity (thus it is
join-preserving).
\end{example}

\begin{proposition}
  \label{prop:Quan-prod}
  The category $\Quan$ has small products.
\end{proposition}
\begin{proof}
  If $\famb{Q_j}{j\in J}$ is a family of objects in $\Quan$, we show
  that $Q\defeq\prod_{j \in J}Q_j$ (see Example~\ref{ex:quantale})
  with the obvious projections is a product in $\Quan$.
  In fact, $Q:\Quan$, by Proposition~\ref{prop:cont_quantales}, and the
  projections $\pi_j:Q\to Q_j$ are in $\Quan$, because they are
  join-preserving and strict-monoidal.

  If $(f_j:Q'\to Q_j \mid j\in J)$ is a family of arrows in $\Quan$, then
  $f\defeq\pair{f_j \mid j\in J}:Q'\to Q$ is an arrow in $\Quan$.
  In fact, $f$ is Scott continuous, because $Q$ is a product in $\CL$,
  and $f$ is lax-unital, \ie, $f(\qI')=(f_j(\qI') \mid j\in
  J)\qle(\qI_j \mid j\in J)$, because all $f_j$ are lax-unital.
\end{proof}

\section{Quantale-valued Metric Spaces}
\label{sec:QMS}

In~\cite{lawvere1973metric} Lawvere proposes to view metric spaces as small
$\RQ_+$-enriched categories and shows that several notions and results
regarding metric spaces are instances of notions and results regarding
$\V$-enriched categories, with $\V$ symmetric monoidal closed category
(see~\cite{kelly1982basic}).  For instance, $\RQ_+$-enriched functors
correspond to short (aka non-expansive) maps between metric spaces.
One can replace $\RQ_+$ with another quantale $Q$ and proceed
in analogy with~\cite{lawvere1973metric}, but care is needed when
$Q$ is not commutative.  Instead we define categories of
quantale-valued metric spaces parametric in a sub-category $\A$ of
$\Quan$, where objects are small $Q$-enriched categories for
$Q\in\A$ and arrows can be more general than short maps.
The rational for restricting to $\Quan$ is to have a forgetful
functor to $\Top$.  In fact, if one allows non-continuous quantales and
lax-unital monotonic maps that are not Scott continuous, then one has
only a forgetful functor to $\Po$.

There are some proof obligations to claim that $\MS_\A$ defined below
is a $\Po$-enriched category.  Therefore, the definition is followed
by a proof of such obligations, which make clear why it is
essential for arrows in $\Quan$ to be lax-unital monotonic maps.
\begin{definition}\label{def:MS-A}\label{def:Q-MS}
  Given a sub-category $\A$ of $\Quan$, the $\Po$-enriched category
  $\MS_\A$ is given by:
  \begin{itemize}
  \item objects are triples $(X,d,Q)$ with $X\in\Set$, $Q\in\A$ and
    $d:X^2\to Q$ such that $d(x,y)\qT d(y,z)\qle d(x,z)$ and $\qI\qle
    d(x,x)$.
    We call $d$ a $Q$-metric, and the \textbf{$d$-preorder} on
    $X$ is $x\leq_d y\defiff\qI\qle d(x,y)$.

  \item arrows from $(X,d,Q)$ to $(X',d',Q')$ are $f:X\to X'$
    such that $\forall x,y\in X.f'(d(x,y))\qle' d'(f(x),f(y))$ for
    some $f':\A(Q,Q')$, called a realizer of $f$, and
    the hom-set preorder is $f_1\leq f_2\defiff\forall x \in
    X.f_1(x)\leq_{d'}f_2(x)$.
  \end{itemize}
  The $\Po$-enriched forgetful
  functor $U:\MS_\A\to\Po$ maps $(X,d,Q)$ to $(X,\leq_q)$ and is the
  identity on hom-sets.
  We write $\MS$ for $\MS_\Quan$.
  An arrow $f:\MS((X,d,Q),(X',d',Q'))$ is an
  \textbf{isometry} when $Q=Q'$ and $\forall x,y \in
  X.d(x,y)=d'(f(x),f(y))$.
\end{definition}
\begin{proof}
  Since $\MS_\A$ is a sub-category of $\MS$ such that the preorder on
  $\MS_\A(X_1,X_2)$ is the restriction of that on $\MS(X_1,X_2)$, it
  suffices to prove the properties for $\MS$.

  First, we define the forgetful functor $U:\MS\to\Po$, by
  proving that $f:\MS((X_1,d_1,Q_1),(X_2,d_2,Q_2))$ implies
  $f:\Po((X_1,\leq_{d_1}),(X_2,\leq_{d_2}))$, \ie, that $f$ is
  monotonic, which means $x\leq_{d_1}y\implies f(x)\leq_{d_2}f(y)$, or
  equivalently $\qI_1\qle_1 d_1(x,y) \implies\qI_2\qle_2
  d_2(f(x),f(y))$ (by definition of $\leq_d$).
  Assume $\qI_1\qle_1 d_1(x,y)$ and let $f' : \Quan(Q,Q')$ be a realizer
  of $f$, then:
  \begin{itemize}
  \item $\qI_2\qle_2 f'(\qI_1)\qle_2 f'(d_1(x,y))$, because $f'$ is lax-unital and monotonic
  \item $f'(d_1(x,y))\qle_2 d_2(f(x),f(y))$, by definition of arrow in
    $\MS$.
  \end{itemize}
  Since the preorder on the hom-sets of $\MS$ is that induced by the
  preorder on the hom-sets of $\Po$ via the forgetful functor, then
  $\MS$ is $\Po$-enriched and the forgetful functor is $\Po$-enriched,
  too.
\end{proof}

In the rest of the paper we may call the objects $(X,d,Q)$ of $\MS$
metric spaces, rather than quantale-valued metric spaces, and a
$Q$-metric $d:X^2\to Q$ simply a metric.  Of course this conflict with
the standard meaning of metric, since the properties of a
quantale-valued metric are weaker that those of a standard metric:
\begin{itemize}
\item the triangle inequality $d(x,z)\leq d(x,y)+d(y,z)$ is
  retained in the form $d(x,y)\qT d(y,z)\qle d(x,z)$
\item $d(x,y)=0\iff x=y$ is replaced by the weaker property $\qI\qle
  d(x,x)$, which in $\RQ_+$-metric spaces corresponds to
  $d(x,x)=0$, since in $\RQ_+$ one has $\qI=0=\qt$
\item symmetry $d(x,y)=d(y,x)$ is not required (it would amount to
  have a self-dual category)
\item \emph{separation}, {\ie}, $d(x,y)=0\implies x=y$, becomes
  $(\qI\qle d(x,y)\land \qI\qle d(y,x))\implies x=y$ when symmetry is
  not assumed.  The metrics with this property are those such that
  $\leq_d$ is a poset (Section~\ref{sec:Po-enriched} gives an abstract
  definition of separated object in $\Po$-enriched categories).
\end{itemize}
Alternatively, we could have borrowed the terminology from
\cite{Goubault-Larrecq:Non_Hausdorff_topology:2013}, where a metric
$d:X^2\to\RQ_+$ in our sense is called a hemi-metric (see
Fig.~\ref{fig:metric-terminology}), Goubault-Larrecq's terminology is
not universally accepted, but has the advantage of not conflicting
with standard terminology.
\begin{figure}[tb]
  \begin{center}
    \begin{tabular}{|l|l|l|}
      \hline
    \textbf{Our terminology} & \textbf{\cite{Goubault-Larrecq:Non_Hausdorff_topology:2013} terminology}&
    \textbf{Property of} $\boldsymbol{d:X^2\to Q}$
    \\\hline \hline
    metric&hemi-metric&
    $\qI\qle d(x,x)$ and $d(x,y)\qT d(y,z)\qle d(x,z)$
    \\\hline
    symmetric metric&symmetric hemi-metric&
    $d(x,y)=d(y,x)$
    \\\hline
    separated metric&$T_0$ hemi-metric&
    $(\qI\qle d(x,y)\land \qI\qle d(y,x))\implies x=y$
    \\\hline
  \end{tabular}
  \end{center}
  \caption{Alternative terminologies for quantale-valued metrics.}
  \label{fig:metric-terminology}
\end{figure}

\begin{remark}
  The $\Pos$-enriched category $\Quan$ and its sub-categories $\Quan_{lm}$, $\Quan_{sm}$ and $\Quan_J$ (see Fig.~\ref{fig:quantale-cats}) are actually
  dcpo-enriched, \ie, the hom-posets are dcpos and composition is
  Scott continuous.  Moreover, $\Quan$ (and $\Quan_J$) has the
  property that every non-empty subset of a hom-dcpo has a join
  computed pointwise, this property implies that every arrow in $\MS$
  has a biggest realizer.
\end{remark}

We summarize some properties of $\MS_Q$ (\ie, the category of
$Q$-valued metric spaces and short maps), that ignore the
$\Po$-enrichment and are proved in~\cite{kelly1982basic} for a small
complete and co-complete symmetric monoidal closed category in place
of a (not necessarily symmetric) quantale $Q$.
\begin{proposition}
  \label{prop:MetQ_has_small_sums_etc}
  The category $\MS_Q$ has small limits and small colimits.
\end{proposition}
\begin{proof}
  Given a family $\famb{(X_i,d_i,Q)}{i\in I}$ of objects in $\MS_Q$, the
  metric on the product $\prod_{i \in I} X_i$ (computed in $\Set$) is
  $d_\Pi(x,y)=\qM_{i \in I}d_i(x_i,y_i)$, and that on the sum
  $\sum_{i \in I} X_i$ is
  $d_\Sigma((j,x),(j',x'))=d_j(x,x')$ if $j=j'$ else $\bot$.

Given a pair of short maps $f,g:\MS_Q((X,d,Q),(X',d',Q))$, the equalizer
is obtained by taking the equalizer $m:X_e \to X$ in $\Set$,
\ie, $X_e=\setb{x \in X}{f(x)=g(x)}$ and $m$ is the inclusion of
$X_e$ into $X$, and endowing $X_e$ with the restriction of $d$ to
it. Then, $m$ is obviously short.
Dually, the coequalizer is obtained by taking the coequalizer $e : X'
\to X_c$ in $\Set$, \ie, $X_c=X'/\approx$, where $\approx$ is the
smallest equivalence relation on $X'$ including the relation
$\setb{(f(x),g(x))}{x\in X}$ and $e$ is the quotient map $x\mapsto
[x]$, and endowing $X_c$ with the metric $d'_\approx$ given by
$d'_\approx([x],[y]) =\qJ_{x'\in[x],y'\in[y]} d'(x',y')$.
\end{proof}

For $\MS$ and some of its sub-categories $\MS_\A$ we prove weaker
properties, that do not follow from~\cite{kelly1982basic}.
\begin{proposition}
  \label{prop:MS-lim+colim}
  For any sub-category $\A$ of $\Quan$:
  \begin{enumerate}
  \item $\MS_\A$ has equalizers and coequalizers.
  \item If $\A$ has small products computed as in $\Quan$, then
    $\MS_\A$ has small products.
  \end{enumerate}
\end{proposition}
\begin{proof}
  If $f,g:(X,d,Q)\to(X',d',Q')$ are arrows in $\MS_\A$, then
  \begin{itemize}
  \item the equalizer of $(f,g)$ is $m:(X_e,d_e,Q)\to(X,d,Q)$, where
    $m:X_e \to X$ is the equalizer of $(f,g)$ in $\Set$, \ie,
    $X_e=\setb{x \in X}{f(x)=g(x)}$, $m$ is the inclusion of $X_e$
    into $X$, $d_e$ is the restriction of $d$ to $X_e$, and
    the arrow $m$ is realized by the identity $\id_Q$.
  \item
    the coequalizer of $(f,g)$ is $e:(X',d',Q')\to(X_c,d_c,Q')$,
    where $e:X'\to X_c$ is the coequalizer of $(f,g)$ in $\Set$, \ie,
    $X_c=X'/\approx$, where $\approx$ is smallest equivalence relation
    on $X'$ including the relation $\setb{(f(x),g(x))}{x\in X}$, $e$
    is the quotient map $x\mapsto [x]$, $d_c$ is the metric on $X_c$
    such that $d_c([x],[y]) =\qJ_{x'\in[x],y'\in[y]} d'(x',y')$, and
    the arrow $e$ is realized by the identity $\id_{Q'}$.
  \end{itemize}
  If $\A$ has $I$-indexed products computed as in $\Quan$ and
  $\famb{(X_i,d_i,Q_i)}{i\in I}$
  is a family of objects in $\MS_\A$,
  then its product is $(X,d,Q)$, where $X$ is the product $\prod_{i \in
    I} X_i$ in $\Set$, $Q$ is the product $\prod_{i\in I}Q_i$ in $\A$
  (and $\Quan$), $d(x,y)\defeq(d_i(x_i,y_i) \mid i\in I)\in Q$, and the
  projection $\pi_i:(X,d,Q)\to(X_i,d_i.Q_i)$ in $\MS_\A$ is realized
  by $\pi_i:Q\to Q_i$ in $\A$.
  Moreover, if $(f_i:(X',d',Q')\to(X_i,d_i,Q_i) \mid i\in I)$ is a family
  of arrows in $\MS_\A$ and $f'_i:Q'\to Q_i$ is a realizer in $\A$ of
  $f_i$ for each $i\in I$, then $\pair{f_i \mid i\in I}:(X',d',Q')\to
  (X,d,Q)$ is an arrow in $\MS_\A$ realized by $\pair{f'_i \mid i\in
    I}:Q'\to Q$.
\end{proof}
\hideEM{CUT: $\MS$ has finite sums, because $\Quan$ has weak sums,
  namely $(\prod_{i\in n}Q_i)_\bot$, and the mediating arrow from the
  weak sum above can always be chosen to be bottom preserving. The
  proof uses that binary meets in a continuous lattice are Scott
  continuous The sum of $\famb{(X_i,d_i,Q_i)}{i\in n}$ is $(X,d,Q)$,
  where where $X$ is the sum $\sum_{i \in n} X_i$ in $\Set$, $Q$ is
  the weak sum $(\prod_{i\in n}Q_i)_\bot$ in $\Quan$,
  $d((i,x),(j,y))\defeq\bot$ if $i\neq j$ else $(d_k(x,y)$ if $k=i$
  else $\qI_k)$, the realizer for $\iota_i:(X_i,d_i,Q_i)\to(X,d,Q)$ is
  easy.
    Moreover, if $(f_i: (X_i,d_i,Q_i)\to(X',d',Q') \mid i\in n)$ is a
    family of arrows in $\MS$ and $f'_i:Q_i\to Q'$ is a realizer of
    $f_i$, then $[f_i \mid i\in n]:(X,d,Q)\to(X',d',Q')$ is realized by
    $f':Q\to Q'$ such that $f'(\bot)=\bot'$ and $f'(q_i \mid i\in
    n)=\qM_{i\in n} f'_i(q_i)$.}


We now define a forgetful functor from $\MS$ to $\Top$, thereby
generalizing the open ball topology induced by a standard metric.
This is possible because the definition of $\MS$ uses the category
$\Quan$, instead of a larger category of quantales.
In general, to an object $(X,d,Q)\in\MS$ one can associate two
topologies on $X$.  Moreover, when $Q$ is $\omega$-continuous---a
restriction desirable from a computational viewpoint
(see~\cite{smyth1977effectively})---convergence in these topologies
can be defined in terms of sequences.
\begin{definition}
  \label{def:open_ball_topology}
  Given $(X,d,Q):\MS$, the \textbf{open ball}
  with center $x\in X$ and radius $\delta\ll\qI$ is
  $B(x,\delta)\defeq\setb{y \in X}{\delta\ll d(x,y)}$.
  The \textbf{open ball topology} $\tau_d$ is the topology
  generated by the family of open balls.
  There is also a \textbf{dual open ball} $B^o(x,\delta)\defeq\setb{y
  \in X}{\delta\ll d(y,x)}$, and the corresponding \textbf{dual open
  ball topology} $\tau_d^o$.
\end{definition}
\begin{remark}
  The definition of open ball uses the way-below relation $\ll$, which
  is well-behaved only in continuous quantales.
  If $d$ is symmetric (\ie, $d(x,y)=d(y,x)$), then the two notions of
  open ball agree.
  If the quantale $Q$ is commutative,
  then one can define the dual metric space $(X,d^o,Q)$, where
  $d^0(x,y)=d(y,x)$, in this case the dual open ball topology
  $\tau_d^o$ coincides with the open ball topology $\tau_{d^o}$ for
  the dual metric $d^o$.
\end{remark}
We focus on open balls, but the results hold \emph{mutatis mutandis}
also for the dual notion.
The following lemma allows to prove that open balls form a \emph{base}
for $\tau_d$, \ie, every open in $\tau_d$ is a union of open balls.
\begin{lemma}\label{lemma:ob:cq}
Open balls satisfy the following properties:
\begin{enumerate}
\item\label{ob:cq:1}
  $x\in B(x,\delta)$.
\item\label{ob:cq:2}
  $\delta\qle\delta'\implies B(x,\delta')\subseteq B(x,\delta)$.
\item\label{ob:cq:3}
  $y\in B(x,\delta)\implies\exists\delta'\ll\qI. B(y,\delta')\subseteq
  B(x,\delta)$.
\item\label{ob:cq:4}
  $y\in B(x_1,\delta_1) \cap B(x_2,\delta_2)\implies \exists
  \delta'\ll \qI.\ B(y,\delta') \subseteq B(x_1,\delta_1) \cap
  B(x_2,\delta_2)$.
\end{enumerate}
\end{lemma}
\begin{proof}
  The proofs are quite straightforward. So, for each property, we give
  only a hint.
\begin{enumerate}
\item Follows from $\delta\ll \qI\qle d(x,x)$.
\item Follows from
  $\delta\qle\delta'\ll d(x,y)\implies\delta\ll d(x,y)$.
\item $y\in B(x,\delta)$ is equivalent to $\delta\ll d(x,y)$. Thus, by
  Lemma~\ref{lemma:inter:cq}, $\delta\ll d(x,y)\qT\delta'$ for some
  $\delta'\ll \qI$.
  Moreover, $B(y,\delta')\subseteq B(x,\delta)$ is equivalent to
  $\delta'\ll d(y,z)\implies \delta\ll d(x,z)$.
  If $\delta'\ll d(y,z)$, then $\delta\ll d(x,y)\qT\delta'\qle
  d(x,y)\qT d(y,z)\qle d(x,z)$, which implies (by
  Proposition~\ref{thm:wb:prop}) $\delta\ll d(x,z)$.
 
\item By item~\ref{ob:cq:3}, $y\in B(x_i,\delta_i)$ implies
  $B(y,\delta'_i)\subseteq B(x,\delta_i)$ for some $\delta'_i\ll\qI$.
  Let $\delta'=\delta'_1\qj\delta'_2$, then
  $\delta'_i\qle\delta'\ll\qI$ (by
  Proposition~\ref{thm:wb:prop}). Thus,
  $B(y,\delta')\subseteq B(y,\delta'_i)\subseteq B(x,\delta_i)$ (by
  item~\ref{ob:cq:2}).
\end{enumerate}
\end{proof}
\begin{proposition}\label{prop:e-d:open}
If $(X,d,Q):\MS$ and $O\subseteq X$, then $O\in\tau_d\iff \forall x
\in O.\exists\delta\ll\qI.B(x,\delta)\subseteq O$.
Moreover, the specialization preorder $\leq_{\tau_d}$ for the open
ball topology $\tau_d$ is $\leq_d$ and the specialization preorder
$\leq_{\tau_d^o}$ is $\leq_d^o$, \ie, the dual of $\leq_d$.
\end{proposition}
\begin{proof}
  The $\Leftarrow$-direction is easy, since the RHS
  says that $O$ is a union of open balls, thus $O\in\tau_d$.

  For the $\Rightarrow$-direction, note that every $O\in\tau_d$ is a
  union of finite intersections of open balls. Therefore, it suffices
  to prove that, for any finite sequence $\famb{B_i}{i \in n}$ of open
  balls $O=\cap_{i \in n} B_i$ satisfies the RHS, which we prove by
  induction on $n$:
  \begin{itemize}
  \item Base case $0$: We note that $\cap \emptyset = X$ and
    $X=B(x,\bot)$ for any $x\in X$. Thus, $\cap \emptyset$ satisfies
    the RHS by choosing $\delta = \bot$.
  
  \item Inductive step $n+1$: by induction hypothesis $O=\cap_{i \in
    n} B_i$ satisfies the RHS. Thus, for any $x\in
    O\cap B_n$, we have $B(x,\delta)\subseteq O$ for some
    $\delta\ll\qI$. In particular, $x\in B(x,\delta)\cap B_n$.
    Therefore, by item~\ref{ob:cq:4} of Lemma~\ref{lemma:ob:cq},
    there exists $\delta'\ll\qI$ such that $B(x,\delta')\subseteq
    B(x,\delta)\cap B_n\subseteq O\cap B_n$.
  \end{itemize}
  We prove that $\leq_{\tau_d}=\leq_d$ by a sequence of equivalences:

  \begin{eqnarray*}
    & & x\leq_{\tau_d}y \\
    (\text{by definition of specialization
    preorder})    & \defiff& (\forall O\in\tau_d. \ x\in O\implies y\in
                             O)\\
    (x\in O \iff\exists\delta\ll\qI. \ B(x,\delta)\subseteq O) &  \iff
      & (\forall\delta\ll\qI. \ y\in B(x,\delta))\\
    (\text{by definition
    of $B(x,\delta)$}) & \iff & (\forall\delta\ll\qI. \ \delta\ll
                                d(x,y))\\
    (\text{$Q$ is
    continuous, thus $\qJ(\qwb\qI)=\qI$}) & \iff & \qI\qle d(x,y) \\
    & \defiff & x\leq_d y.\\
  \end{eqnarray*}
  The proof of $x\leq_{\tau_d^o}y\iff y\leq_d x$ is similar, the only
  difference is $y\in B^o(x,\delta)\iff\delta\ll d(y,x)$.
\end{proof}

We characterize continuous maps with respect to open ball
topologies in terms of metrics (using the familiar
\emph{epsilon-delta} formulation) and prove that arrows in $\MS$
are the \emph{uniformly continuous} maps.
\begin{theorem}\label{thm:e-d:continuity}
  If $(X_i,d_i,Q_i):\MS$ for $i=1,2$ and $f:X_1\to X_2$, then
  $$f:\Top((X_1,\tau_{d_1}),(X_2,\tau_{d_2}))\iff
  \forall x \in X_1.\forall\epsilon\ll
  \qI_2.\exists\delta\ll\qI_1.f(B(x,\delta))\subseteq
  B(f(x),\epsilon).$$
\end{theorem}
\begin{proof}
  For the $\Rightarrow$-direction, assume the LHS, \ie, $\forall
  O\in\tau_{d_2}.\inv{f}(O)\in\tau_{d_1}$.  Let $O\in\tau_{d_2}$ be
  the open ball $B(f(x),\epsilon)$. Then, by
  Proposition~\ref{prop:e-d:open}, there exists a $\delta\ll\qI_1$
  such that $B(x,\delta)\subseteq\inv{f}(O)$, which is equivalent to
  $f(B(x,\delta))\subseteq O=B(f(x),\epsilon)$.

  For the $\Leftarrow$-direction, we assume the RHS and
  $O\in\tau_{d_2}$ and prove that $O'=\inv{f}(O)\in\tau_{d_1}$, or
  equivalently (by Lemma~\ref{lemma:ob:cq}) for any $x\in O'$,
  there exists $\delta\ll\qI$ such that $B(x,\delta)\subseteq O'$. If
  $x\in O'$, then $f(x)\in O$. Hence, by Proposition~\ref{prop:e-d:open},
  $B(f(x),\epsilon)\subseteq O$ for some $\epsilon\ll\qI_2$, and by
  the RHS, there exists a $\delta\ll\qI_1$ such that
  $f(B(x,\delta))\subseteq B(f(x),\epsilon)\subseteq O$, which implies
  $B(x,\delta)\subseteq O'$.
\end{proof}

The following theorem shows that the arrows in
$\MS((X,d,Q),(X',d',Q'))$ are the \emph{uniformly continuous}
maps.  Therefore the mapping of $f:\MS((X,d,Q),(X',d',Q'))$ to
$f:\Top((X,\tau_d),(X',\tau_{d'}))$ defines a ($\Po$-enriched)
faithful functor from $\MS$ to $\Top$.
\begin{theorem}\label{thm:e-d:uniformity}
  If $(X_i,d_i,Q_i):\MS$ for $i=1,2$ and $f:X_1\to X_2$, then
  $$f:\MS((X_1,d_1,Q_1),(X_2,d_2,Q_2))\iff
  \forall\epsilon\ll
  \qI_2.\exists\delta\ll\qI_1.\forall x \in X_1.f(B(x,\delta))\subseteq
  B(f(x),\epsilon).$$
\end{theorem}
\begin{proof}
  For the $\Rightarrow$-direction, assume $\epsilon\ll\qI_2$ and the
  LHS, \ie, $\forall x,y:X_1.f'(d_1(x,y))\qle d_2(f(x),f(y))$ for some
  $f':\Quan(Q_1,Q_2)$.
  Since $f'$ is Scott continuous, $\epsilon\ll\qI_2\qle f'(\qI_1)$ and
  $\qwa\epsilon$ is Scott open in the continuous lattice $Q_2$, then  
  $f'(\qwa\delta)\subseteq\qwa\epsilon$ for some $\delta\ll\qI_1$.
  Therefore, $\delta\ll d_1(x,y)$ implies $\epsilon\ll
  f'(d_1(x,y))\qle d_2(f(x),f(y))$ for every $x,y\in X_1$, or
  equivalently $f(B(x,\delta)\subseteq B(f(x),\epsilon)$ for every
  $x\in X_1$.

  For the $\Leftarrow$-direction, we assume the RHS and
  we must define a realizer $f':\Quan(Q_1,Q_2)$ for $f$.
  Consider the subset $F'$ of the continuous lattice $\CL(Q_1,Q_2)$ of
  Scott continuous maps from $Q_1$ to $Q_2$ consisting of the $g'$
  such that $\forall x,y:X_1.g'(d_1(x,y))\qle d_2(f(x),f(y))$.
  Let $f'\defeq \qJ F'\in\CL(Q_1,Q_2)$, then $f'$ is the maximum of
  $F'$, because $F'$ is closed under joins computed in $\CL(Q_1,Q_2)$.
  To prove that $f'$ is a realizer of $f$, it suffices to prove
  $\qI_2\qle f'(\qI_1)$, or equivalently
  $\forall\epsilon\ll\qI_2.\epsilon\qle f'(\qI_1)$ (since $Q_2$ is
  continuous).
  Given $\epsilon\ll\qI_2$ there exists (by the RHS) a
  $\delta\ll\qI_1$ such that $\forall x \in
  X_1.f(B(x,\delta))\subseteq B(f(x),\epsilon)$, or equivalently
  $\delta\ll d_1(x,y)\implies\epsilon\ll d_2(f(x),f(y))$.  Hence, the
  step function $[\delta,\epsilon](q)\defeq\epsilon$ if $\delta\ll q$
  else $\bot_2$ is in $F'$, and we can conclude
  $\epsilon=[\delta,\epsilon](\qI_1)\qle f'(\qI_1)$.
\end{proof}

\hideEM{HIDE: the $\Po$-enriched forgetful functor from $\MS$ to
  $\Po$ factors through $\Top$.  TODO: what do these functors
  preserve? Is it useful for the rest of the paper?}

\subsection{Extensive Metrization}
\label{subsec:Extensive_Metrization}

We prove that every topology on a set $X$ is the open ball topology
$\tau_d$ of some metric space $( X, d, Q) : \MS$, by exploiting a
similar result in~\cite{Flagg:Quantales_continuity_spaces:1997} that
every topology is induced by a generalized metric defined using
prime-continuous quantales and the $\lll$ relation.

A description of Flagg's construction compatible with our framework is
given in~\cite[Example~11]{Cook_Weiss:quantales_Lip:2022}, which uses
the prime-algebraic lattices $\Omega(S) = \DM(\PM_f(S),\subseteq)$,
where $S$ is a set.  In fact, $\Omega(S)$ is a prime-algebraic locale
(see properties of $\DM(P)$ in Example~\ref{ex:continuous-quantale}).
As such, it suffices to prove the following result.
\begin{lemma}
  \label{lemma:CDL_WB_Total_Top}
  If $(Q, \qT, \qI)$ is a prime-continuous quantale and $(X, d, Q) :
  \MS$, let $\tau_d^T$ be the topology generated by the open
  balls $B_T(x,\delta) \defeq \setb{y \in X}{\delta \lll d(x,y)}$, that
  use $\lll$ instead of $\ll$ (in
  Definition~\ref{def:open_ball_topology}), then $\tau_d^T = \tau_d$.
\end{lemma}
\begin{proof}
  We first prove that $\tau_d^T \subseteq \tau_d$. It suffices to show
  that $\forall x \in X. \forall a \lll \qI. \exists b \ll \qI.  B(x,
  b) \subseteq B_T(x,a)$.  Assume that $a \lll \qI$, from the
  interpolation property of the totally-below relation
  (Lemma~\ref{lemma:inter:cl}), we obtain $a \lll b \lll \qI$ for some
  $b\in Q$.  Therefore, for every $y \in B(x,b)$ we derive
  $b \ll d( x, y) \implies a \lll d( x, y) \implies y \in B_T( x, a)$.

  To prove $\tau_d \subseteq \tau_d^T$, it suffices to show that:
  \begin{equation}
   \label{eq:cap_B_T-B}
    \forall x \in X . \forall e \ll \qI . \exists a_1, \ldots, a_n
    \lll \qI. \quad \bigcap_{i=1}^n B_T( x, a_i) \subseteq B( x, e).
  \end{equation}

  Since $Q$ is prime-continuous, we have $\qI = \qJ A$ when $A \defeq
  \setb{a \in Q}{ a \lll u}$. Let $A_\qj \defeq \setb{\qJ
    F}{F\subseteq_f A}$ be the closure of $A$ under finite joins.
  Since $A_\qj$ is a directed subset of $Q$ and $\qI = \qJ A_\qj$, for
  any $e \ll u$ there exists $F \defeq \setf{a_1, \ldots, a_n}
  \subseteq_f A$ such that $e \qle \qJ F$. This
  proves~\eqref{eq:cap_B_T-B}, because of the following chain of
  implications: $\forall a\in F.a\lll d(x,y)\implies\forall a\in
  F.a\ll d(x,y)\implies \qJ F\ll d(x,y)\implies e\ll d(x,y)$.
\end{proof}

\begin{theorem}
  \label{thm:extensive_metrizability}
  Every topology $\tau$ on $X$ coincides with the open ball topology $\tau_d$
  of a metric space $(X, d, Q) : \MS$.
\end{theorem}
\begin{proof}
  From Lemma~\ref{lemma:CDL_WB_Total_Top} and~\cite[page 45]{CookW21}.
  More precisely, $Q$ is the prime-algebraic locale $\Omega(S)$, where
  $S$ is the set $\tau$ of open subsets of $X$, and
  $d(x,y)=\PM_f(\setb{O\in\tau}{x\in O\implies y\in O})$.
\end{proof}

\subsection{Imprecision and Robustness}
\label{sec:part1}

We extend the notions of imprecision and robustness
in~\cite{Moggi_Farjudian_Duracz_Taha:Reachability_Hybrid:2018,MoggiFT-ICTCS-2019}
from standard metric spaces to objects $(X,d,Q)$ in $\MS$.
Since $d$ may fail to be symmetric, we must consider the ``direction''
along which the distance is measured.  For example, in the presence
of imprecision, two subsets are indistinguishable when they have the
same closure in the dual open ball topology $\tau_d^o$, rather than in
the topology $\tau_d$ (Proposition~\ref{prop:spec-robust-top}).  This
difference cannot be appreciated when $d$ is symmetric, because the
two topologies coincide.

The following result characterizes the closure of a subset $A$ with
respect to the open ball topology as the set of points \emph{from}
which one can reach a point in $A$ within any arbitrarily small
distance (and a similar result holds for the dual open ball topology).
\begin{lemma}\label{lemma:e-d:closed}
  If $(X,d,Q):\MS$ and $A\subseteq X$, then the closure of $A$ in
  $(X,\tau_d)$ is given by:
  \begin{equation*}
    \cl{A}=\setb{y \in X}{\forall\delta\ll
    \qI.\exists x \in A.\delta\ll d(y,x)}.
  \end{equation*}
\end{lemma}
\begin{proof}
  Let $F \defeq \setb{y \in X}{\forall\delta\ll \qI.\exists x \in
    A.\delta\ll d(y,x)}$. To prove the lemma, we must show that $F$ is
  the smallest closed set that contains $A$, or equivalently, $X
  \setminus F$ is the largest open set that does not intersect
  $A$. For any $z \in X$, we have:
  \begin{eqnarray*}
    & & z \in X \setminus F \\
    (\text{By definition of $F$})& \iff & \exists\delta\ll\qI.\forall x \in A.\delta\not\ll
                                        d(z,x) \\
    (\text{By definition of $B( z, \delta)$}) & \iff &
                                                       \exists\delta\ll\qI.B(z,\delta)\cap A=\emptyset ,    
  \end{eqnarray*}
  which proves the lemma.
\end{proof}

\begin{definition}
  The notions in~\cite[Definition~1]{MoggiFT-ICTCS-2019} generalize as
  follows to $(X,d,Q)\in\MS$:
\begin{enumerate}
\item
  $B_R(A,\delta)\defeq \setb{y \in X}{\exists x \in A.\delta\ll
    d(x,y)}= \cup_{x\in A}B(x,\delta)$ is the set of points in $X$
  belonging to $A\subseteq X$ with precision greater than
  $\delta\ll\qI$.\footnote{The terminology used
    in~\cite{MoggiFT-ICTCS-2019} is ``with imprecision less than
    $\delta$''.}

\item $A_\delta\defeq\cl{B_R(A,\delta)}^o$ is the
  \textbf{fattening} of $A\subseteq X$ by $\delta\ll\qI$,
  where $\cl{Y}^o$ is the closure of $Y$ in $\tau_d^o$ (see
  Lemma~\ref{lemma:e-d:closed}).
\end{enumerate}
\end{definition}

\begin{proposition}
  \label{prop:BR_A_delta_properties}
The subsets $B_R(A,\delta)$ have the following properties:
\begin{enumerate}
\item \label{prop:BR_A_delta_properties:mon} 
  $A\subseteq B_R(A,\delta)\subseteq B_R(A',\delta')$ when
  $A\subseteq A'\subseteq X$ and $\delta'\qle\delta\ll\qI$.
    
\item \label{prop:BR_A_delta_properties:tensor} 
  $B_R(B_R(A,\delta_1),\delta_2)\subseteq B_R(A,\delta)$ when
  $\delta_1,\delta_2\ll\qI$ and
  $\delta\ll\delta_1\qT\delta_2[\qle\delta_i]$.

\item \label{prop:BR_A_delta_properties:cl} 
  $\cl{A}^o=\cap_{\delta\ll\qI}B_R(A,\delta)$ for every $A\subseteq X$.

\item \label{prop:BR_A_delta_properties:cl-eq} 
  $B_R(\cl{A}^o,\delta)=B_R(A,\delta)$ for every $A\subseteq X$
  and $\delta\ll\qI$, \ie, $A$ and $\cl{A}^o$ are indistinguishable in
  the presence of imprecision.

\item \label{prop:BR_A_delta_properties:falt} 
  $B_R(A,\delta)\subseteq A_\delta\subseteq B_R(A,\delta')$ when
  $A\subseteq X$ and $\delta'\ll\delta\ll\qI$.
\end{enumerate}
\end{proposition}
\begin{proof}
  For each property we give a proof hint.
\begin{enumerate}
\item Follows easily from the definition of $B_R(A,\delta)$.
    
\item Under the assumption $\delta_1,\delta_2\ll\qI$, one has
  $\delta_1\qT\delta_2\qle\delta_1\qT\qI=\delta_1\ll\qI$ and
  $\delta_1\qT\delta_2\qle\qI\qT\delta_2=\delta_2\ll\qI$.
  If $z\in B_R(B_R(A,\delta_1),\delta_2)$, then $\delta_2\ll d(y,z)$
  for some $y\in B_R(A,\delta_1)$. Hence, $\delta_2\ll d(y,z)$ and
  $\delta_1\ll d(x,y)$ for some $x\in A$, thus
  $\delta\ll\delta_1\qT\delta_2\qle d(x,y)\qT d(y,z)\qle d(x,z)$.

\item Follows easily from Lemma~\ref{lemma:e-d:closed} and the
  definition of $B_R(A,\delta)$.

\item It suffices to prove the inclusion
  $B_R(\cl{A}^o,\delta)\subseteq B_R(A,\delta)$.  If $z\in
  B_R(\cl{A}^o,\delta)$, then $\delta\ll d(y,z)$ for some
  $y\in\cl{A}^o$.  Choose $\epsilon\ll\qI$ such that
  $\delta\ll\epsilon\qT d(y,z)$ and $x\in A$ such that $\epsilon\ll
  d(x,y)$, then $\delta\ll\epsilon\qT d(y,z)\qle d(x,y)\qT d(y,z)\qle
  d(x,z)$.

\item The first inclusion follows from the definition of
  $A_\delta$.
  Since $A_\delta\subseteq
  \cap_{\epsilon\ll\qI}B_R(B_R(A,\delta),\epsilon)$, for the second
  inclusion it suffices to choose $\epsilon\ll\qI$ such that
  $\delta'\ll\delta\qT\epsilon$, then
  $B_R(B_R(A,\delta),\epsilon)\subseteq B_R(A,\delta')$.
\end{enumerate}
\end{proof}

\begin{figure}[h]
  \centering
  \scalebox{0.35}[0.35]{\includegraphics{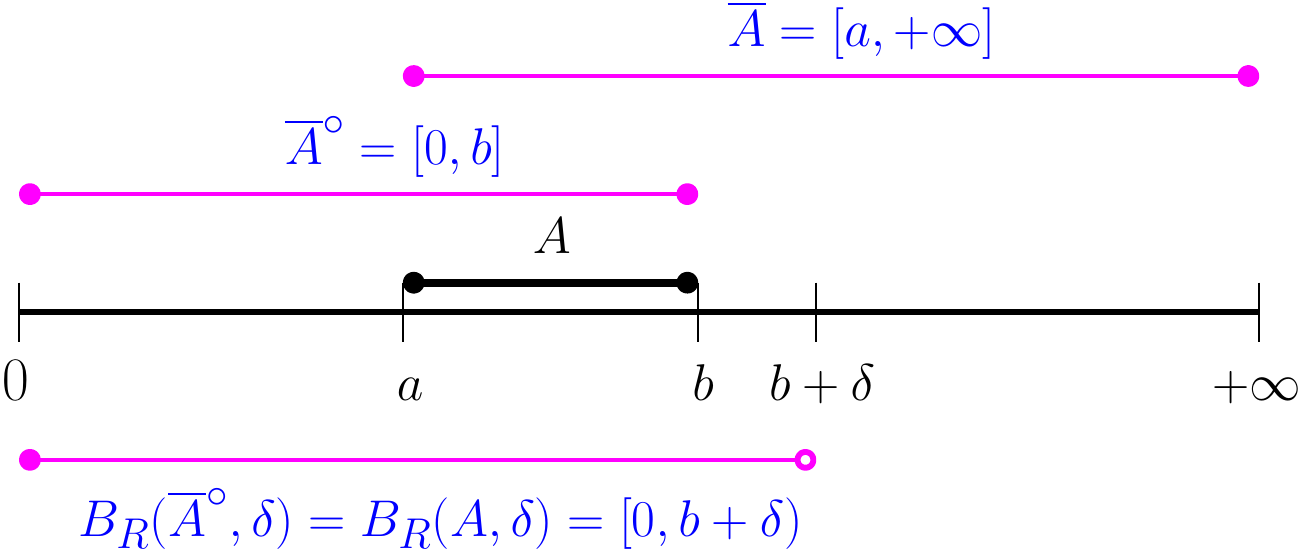}}
  \caption{Graphic recast of
    Example~\ref{example:non_symmetric_R_Plus}.}
  \label{fig:R_Plus}
\end{figure}

\begin{example}
  \label{example:non_symmetric_R_Plus}
  Consider $(X,d.Q)$, where $Q = X = \RQ_+$ and
  $d(x,y)\defeq y-x$ if $x \leq y$ else $0$.  If $A=[a,b]$ and
  $\delta \in (0,+\infty)$, then $\cl{A}= [a,+\infty]$, $\cl{A}^\circ
  = [0,b]$, and $B_R(\cl{A}^o,\delta)=B_R(A,\delta) = [0,b+\delta)$,
    as depicted in Fig.~\ref{fig:R_Plus}.
\end{example}

Also the definition of robust topology in~\cite[Definition
  A.1]{Moggi_Farjudian_Duracz_Taha:Reachability_Hybrid:2018}
generalizes to $(X,d,Q)\in\MS$.
We define this topology on $\PM(X)$, rather than on the set of closed
subsets in the topology $\tau_d^o$, since the restriction to the set
of closed subsets amounts to replacing a topological space with a
\emph{separated} space (see Section~\ref{sec:Po-enriched}).

\begin{definition}\label{def:top:UR}
  Given $(X,d,Q)\in\MS$, the \textbf{robust topology} $\tau_{d,R}$ on
  $\PM(X)$ is given by:
$$U\in\tau_{d,R}\defiff \forall A\in
  U.\exists\delta\ll\qI.\PM(B_R(A,\delta))\subseteq U.$$ 
\end{definition}
In general, the robust topology $\tau_{d,R}$ is not determined by the
open ball topology $\tau_d$.
\begin{example}\label{ex:tauR_counterexample}
  Consider two $\RQ_+$-metrics on $X\defeq\setf{2^{-n} \mid n\in\omega}$:
  the Euclidean metric $d(x,y)=|x-y|$ and the discrete metric
  $d'(x,y)=\mbox{$0$ if $x=y$ else $1$}$. Their open ball topologies
  are the same, namely $\tau_d=\tau_{d'}$ is the discrete topology on
  $X$, but their robust topologies differ, namely
  $\tau_{d,R}\subset\tau_{d',R}$:
  \begin{itemize}
  \item $\tau_{d',R}$ is Alexandrov topology for $\PM(X)$ ordered by
    reverse inclusion, because $B'_R(A,\delta)=A$ when $A\subseteq X$
    and $0<\delta<1$, and $\tau_{d,R}\subseteq\tau_{d',R}$, because
    $B'_R(A,\delta)\subseteq B_R(A,\delta)$ for any $A\subseteq X$ and
    $0<\delta$.
  \item The inclusion is strict, because if
    $A\defeq\setf{2^{-2n} \mid n\in\omega}\subseteq X$, then
    $U\defeq\PM(A)\in\tau_{d',R}$, but $U\notin\tau_{d,R}$.  In fact,
    $A\subset B_R(A,\delta)$ for any $\delta>0$, because $\forall
    n\in\omega.2^{-n}<\delta\implies 2^{-n}\in B_R(A,\delta)$.
  \end{itemize}
\end{example}
  
Finally, we characterize the specialization preorder
$\leq_{\tau_{d,R}}$ for the robust topology $\tau_{d,R}$ As a
consequence, we have that two subsets are indistinguishable in
$\tau_{d,R}$ exactly when they have the same closure in $\tau_d^o$.
\begin{proposition}
  \label{prop:spec-robust-top}
  Given $(X,d,Q)\in\MS$ and $A,B\subseteq X$, we have
  $A\leq_{\tau_{d,R}} B \iff B \subseteq \cl{A}^o$.
\end{proposition}
\begin{proof}
  For the $\Leftarrow$-direction, we prove that $A\in
  U\in\tau_{d,R}$ implies $B\in U$.  By definition of
  $\tau_{d,R}$, we have $\PM(B_R(A,\delta)) \subseteq U$, for some
  $\delta \ll \qI$.  By hypothesis and
  Proposition~\ref{prop:BR_A_delta_properties} we get $B \subseteq
  \cl{A}^o \subseteq B_R(\cl{A}^o,\delta) = B_R(A,\delta)$. Thus, $B
  \in U$, as required.

  For the $\Rightarrow$-direction, we prove the logically equivalent
  $B \not\subseteq \cl{A}^o \implies(\exists U\in\tau_{d,R}.A\in
  U\land B\notin U)$.
  If $B \not\subseteq \cl{A}^o$, then there is $x\in B$ such that
  $x\notin \cl{A}^o$.
  By Proposition~\ref{prop:BR_A_delta_properties}, we have $\cl{A}^o =
  \bigcap_{\delta\ll\qI} B_R(A,\delta)$. Thus, there is $\delta \ll
  \qI$ such that $x \notin B_R(A,\delta)$, and consequently, $x\notin
  B_R(A,\delta')$ for every $\delta'$ such that $\delta\ll \delta'\ll
  \qI$.
  We define $U\in\tau_{d,R}$ such that $A\in U$ and $B\not\in U$.  Let
  $U=\bigcup_{\delta\ll\delta'\ll\qI} \PM(B_R(A,\delta'))$. Clearly,
  $A\in U$, because by the interpolation property
  (Lemma~\ref{lemma:inter:cl}) there is at least one $\delta'$ such
  that $\delta\ll \delta'\ll \qI$, and $B\not\in U$, since $x\notin
  B_R(A,\delta')$ for every $\delta'$ such that $\delta\ll \delta'\ll
  \qI$.
  We now prove that $U\in\tau_{d,R}$, \ie, for every $A'\in U$ there
  exists $\delta_2\ll\qI$ such that $\PM(B_R(A',\delta_2))\subseteq
  U$.
  $A'\in U$ means that $A'\subseteq B_R(A,\delta_1)$ for some
  $\delta\ll\delta_1\ll\qI$.
  By Lemma~\ref{lemma:inter:cl}~and~\ref{lemma:inter:cq}, there are
  $\delta',\delta_2\ll\qI$ such that $\delta \ll \delta' \ll
  \delta_1\qT\delta_2 \ll \qI$.
  Hence, by Proposition~\ref{prop:BR_A_delta_properties} we get
  $B_R(A',\delta_2) \subseteq B_R(B_R(A,\delta_1),\delta_2) \subseteq
  B_R(A,\delta') $.  Therefore, we have $\PM(B_R(A',\delta_2))
  \subseteq \PM(B_R(A,\delta')) \subseteq U$.
\end{proof}


\section{Separation in Preorder-enriched Categories}
\label{sec:Po-enriched}

Preorders, topological spaces and similar structures have a notion of
\emph{indistinguishability} between elements.  Informally, in such
structures, \emph{separation} can be understood as the property
requiring that indistinguishable elements are equal.
We define and study this notion in the setting of $\Po$-enriched
categories. In particular, we show that the definition of separation
in this abstract setting subsumes several \emph{concrete} definitions
within specific categories, including the category $\MS$.
\begin{definition}[Separation]
  Given a $\Po$-enriched category $\A$, we say that:
  \begin{enumerate}
  \item $f,g\in\A(X,Y)$ are \textbf{equivalent} (notation
    $f\sim g$) $\defiff (f\leq g) \land (g\leq f)$.
  \item the hom-preorder $\A(X,Y)$ is separated $\defiff$ it is a
    poset.
  \item an object $Y\in\A$ is separated $\defiff\A(X,Y)$
    is separated for every $X\in\A$.
  \item $\A$ is separated $\defiff Y$ is separated for every $Y\in\A$,
    \ie, $\A$ is $\Pos$-enriched.
  \end{enumerate}
\end{definition}
\begin{remark}
  The definition of ``$\A(X,Y)$ is separated'' can be recast in terms
  of equivalence, \ie, $f\sim g\implies f=g$, for every $f,g:\A(X,Y)$.
  There is a similar recast also for the definition of ``$\A$ is
  separated'', \ie, $f\sim g\implies f=g$, for every pair $(f,g)$ of
  parallel arrows in $\A$.
  In some $\Po$-enriched categories, separated objects have a
  direct characterization that does not refer to arrows:
\begin{enumerate}
\item in $\Po$, separated objects are posets.
\item In $\Top$, separated objects are $T_0$-spaces.
\item In $\MS$, separated objects are those $(X,d,Q)$ in which
  the preorder $\leq_d$ is a poset (see Section~\ref{sec:QMS}).
\end{enumerate}
\end{remark}

\begin{definition}
  \label{def:s-construction}
  $\s\A$ denotes the full $\Po$-enriched sub-category of $\A$
  consisting of separated objects.
\end{definition}
A $\Po$-enriched category $\A$ is separated exactly when $\s\A=\A$.  A
weaker property (satisfied by $\Po$, $\Top$, and $\MS$) is that
every object is \emph{equivalent} (in the sense of
Definition~\ref{def:po-adj}) to one in $\s\A$.

The following property of the sub-categories $\MS_Q$ extends to any
other sub-category $\MS_\A$ of $\MS$.
\begin{proposition}
  \label{prop:MetQ_every_object_separated}
  Given a quantale $Q\in\Quan$, every object in $\MS_Q$ is equivalent
  to a separated one.
\end{proposition}
\begin{proof}
  Given an object $(X,d,Q)$, let $\sim_d$ be the equivalence on $X$
  induced by the preorder $\leq_d$, \ie, $x\sim_d 
  y\defiff\qI\qle d(x,y)\land \qI\qle d(y,x)$. Let $X_0$ be the
  quotient $X/\sim_d$ and define $d_0:X_0\times X_0 \to Q$ as
  $d_0(\eqc{x},\eqc{y})=d(x,y)$.  Since $x\sim_d x'\land y\sim_d
  y'\implies d(x,y)=d(x',y')$, the metric $d_0$ is well-defined.

  Let $\r:X \to X_0$ be the map such that $\r(x)=\eqc{x}$, which is an
  isometry from $(X,d,Q)$ to $(X_0,d_0,Q)$, because
  $d_0(r(x),r(y))=d(x,y)$.  Since $\r$ is surjective, there is a
  section $\sr:X_0 \to X$, which chooses a representative from each
  equivalence class $\eqc{x}\in X_0$, moreover any section $\sr$ is
  also an isometry from $(X_0,d_0,Q)$ to $(X,d,Q)$, because
  $d(\sr(\eqc{x}),\sr(\eqc{y}))= d(x,y)=d_0(\eqc{x},\eqc{y})$.
  To prove that $(\r,\sr)$ is an equivalence in $\MS_Q$, \ie,
  $\r\circ\sr\sim\id_{X_0}$ and $\sr\circ\r\sim\id_{X}$, where $\sim$
  on $\MS_Q((X,d,Q),(X',d',Q))$ is the pointwise extension of $\sim_{d'}$,
  it suffices to observe that $\r(\sr(\eqc{x}))=\eqc{x}$ and
  $\sr(\r(x))\sim_d x$ for every $x\in X$.
\end{proof}

If every object in $\A$ is equivalent to a separated one, then every
$\Po$-enriched endofunctor on $\A$ can be \emph{transformed} into one
that factors through $\s\A$.  Since we are interested in monads, we
describe this \emph{transformer} directly on the category $\Mon(\A)$ of
$\Po$-enriched monads on $\A$ (see Definition~\ref{def:Mon}).
\begin{definition}\label{def:MonT}
  A basic \textbf{monad transformer} on $\Mon(\A)$ is a pair
  $(\T,\inT)$, where $\T$ is function on the objects of $\Mon(\A)$ and
  $\inT$ is a family of monad maps $\inT_{\hat{M}}$ from $\hat{M}$ to
  $\T\hat{M}$.
\end{definition}
\begin{theorem}\label{thm:endo:transformer}
  If $\famb{\r_X:X\to\R X} {X: \A}$ is a family of arrows in a
  $\Po$-enriched category $\A$ such that:
  \begin{equation}\label{prop:weak:equiv}
    \text{$\R X: \s\A$ and $(\r_X,\sr_X)$ is an equivalence for some $\sr_X:\R X\to X$},
  \end{equation}
  then $(\T,\inT)$ defined below is a monad transformer on
  $\Mon(\A)$:
\begin{itemize}
\item $\T$ is the function mapping $\hat{M}=(M,\eta,-^*)$
  to $\T\hat{M}=(M',\eta',-^{*'})$, where
  \begin{itemize}
  \item $M'X\defeq \R(MX)$
  \item $\eta'_X\defeq \r_{MX}\circ\eta_X:\A(X,M'X)$
  \item if $f:\A(X,M'Y)$, then
    $f^{*'}\defeq\r_{MY}\circ(\sr_{MY}\circ f)^*\circ\sr_{MX}:\A(M'X,M'Y)$.
  \end{itemize}

\item $\inT$ is the family of monad maps such that
  $\inT_{\hat{M},X}\defeq\r_{MX}:\A(MX,M'X)$.
\end{itemize}
Moreover, the definition of $\T$ is independent of the choice of $\sr_X$.
\end{theorem}
\begin{proof}
  All the maps on hom-preorders used in the definition of $\T$ are
  monotonic, thus they preserve $\sim$.  Therefore, when $Y$ is
  separated, to prove that two arrows $f,g\in\A(X,Y)$ defined by
  different monotonic constructions are equal, it suffices to prove
  that they are equivalent (\ie, $f\sim g$).
  For the same reason, if in a monotonic construction, one can replace 
  $\sr_X$ with another $\sr'_X$ such that $(\r_X,\sr'_X)$ is an
  equivalence, the results will be equivalent, because
  $\sr_X\sim\sr'_X$.

  $\T\hat{M}=(M',\eta',-^{*'})$ satisfies the equations
  in Definition~\ref{def:po-monad}, namely:
  \begin{itemize}
  \item $f^{*'}\circ\eta'_X=f:X\to\R(MY)$ when $f:X\to\R(MY)$,
    because:
    \begin{align*}    
      f^{*'}\circ\eta'_X &= \r_{MY}\circ(\sr_{MY}\circ
      f)^*\circ\sr_{MX}\circ\r_{MX}\circ\eta_X \\ &\sim
      \r_{MY}\circ(\sr_{MY}\circ f)^*\circ\eta_X =
      \r_{MY}\circ\sr_{MY}\circ f = f .
    \end{align*} 

  \item $(\eta'_X)^{*'}=\id_{M'X}:\R(MX)\to\R(MX)$, because:
    \begin{align*} 
      {\eta'_X}^{*'} &=
      \r_{MX}\circ(\sr_{MX}\circ\r_{MX}\circ\eta_X)^*\circ\sr_{MX}
      \\ &\sim \r_{MX}\circ\eta_X^*\circ\sr_{MX} =
      \r_{MX}\circ\sr_{MX} = \id_{M'X}.
       \end{align*} 

  \item $g^{*'}\circ f^{*'}=(g^{*'}\circ f)^{*'}:\R(MX)\to\R(MZ)$ when
    $f:X\to\R(MY)$ and $g:Y\to\R(MZ)$, because:
    \begin{align*}
        g^{*'}\circ f^{*'} &= \r_{MZ}\circ(\sr_{MZ}\circ
        g)^*\circ\sr_{MY}\circ \r_{MY}\circ(\sr_{MY}\circ
        f)^*\circ\sr_{MX} \\ &\sim \r_{MZ}\circ(\sr_{MZ}\circ
        g)^*\circ(\sr_{MY}\circ f)^*\circ\sr_{MX} \\ &=
        \r_{MZ}\circ((\sr_{MZ}\circ g)^*\circ\sr_{MY}\circ
        f)^*\circ\sr_{MX} \\ &\sim
        \r_{MZ}\circ(\sr_{MZ}\circ\r_{MZ}\circ (\sr_{MZ}\circ
        g)^*\circ\sr_{MY}\circ f)^*\circ\sr_{MX} \\ &=
        \r_{MZ}\circ(\sr_{MZ}\circ g^{*'}\circ f)^*\circ\sr_{MX} =
        (g^{*'}\circ f)^{*'} .
    \end{align*}
  \end{itemize}
  $\inT_{\hat{M}}$ satisfies equations~(\ref{eq:monad:map}) for a
  monad map from $\hat{M}$ to $\T\hat{M}$, namely:
  \begin{itemize}
  \item $\inT_{\hat{M},X}\circ\eta_X=\eta'_X:X\to\R(MX)$, because:
    \begin{align*} 
      \inT_{\hat{M},X}\circ\eta_X &= \r_{MX}\circ\eta_X = \eta'_X .
      \end{align*} 
  \item $\inT_{\hat{M},Y}\circ f^*=(\inT_{\hat{M},Y}\circ
    f)^{*'}\circ\inT_{\hat{M},X}:MX\to\R(MY)$ when $f:X\to MY$,
    because:
    \begin{align*} \inT_{\hat{M},Y}\circ f^* &= \r_{MY}\circ
    f^* \\ &\sim \r_{MY}\circ(\sr_{MY}\circ\r_{MY}\circ
    f)^*\circ\sr_{MX}\circ\r_{MX} \\ &= (\inT_{\hat{M},Y}\circ
    f)^{*'}\circ\inT_{\hat{M},X} .
    \end{align*} 
  \end{itemize}
\end{proof}
We use the simplest form of monad transformer (\ie, \emph{basic
transformer}) among those in~\cite{JMoggi2010monad}.  However, the
monad transformer described in the theorem can be shown to be a
\emph{monoidal transformer}.
The category $\Mon(\A)$ can be made $\Po$-enriched, which is
needed for defining equivalence.
Although the theorem says that $\inT_{\hat{M},X}$ is part of an
equivalence between $MX$ and $\R(MX)$ in $\A$, this does not imply
that $\inT_{\hat{M}}$ is part of an equivalence between $\hat{M}$ and
$\T\hat{M}$ in $\Mon(\A)$ made $\Po$-enriched.

\section{The Hausdorff-Smyth Monad}
\label{sec:Monads}

In this section we define a $\Po$-enriched monad $\hat{\PM}_S$ on $\MS$,
which \emph{captures} the robust topology, \ie, if
$\PM_S(X,d,Q)=(\PM(X),d_S,Q_S)$, then $\tau_{d_S}$ is the robust
topology $\tau_{d,R}$ given in Definition~\ref{def:top:UR}.
In particular, $\hat{\PM}_S$ is a \emph{lifting} of the powerset monad
$\hat{\PM}$ on $\Set$ along the forgetful faithful functor $U:\MS\to\Set$.
Moreover, we identify sufficient conditions on a sub-category $\A$ of $\Quan$,
which ensures that $\hat{\PM}_S$ restricts to $\MS_\A$.

Since $\hat{\PM}_S$ is $\Po$-enriched, the monad transformer of
Section~\ref{sec:Po-enriched} allows to get a separated version of
$\hat{\PM}_S$, which partitions $\PM(X)$ into equivalence classes with
canonical representatives (see
Proposition~\ref{prop:spec-robust-top}).
\begin{remark}
  The forgetful functor $U:\MS\to\Set$ factors through the forgetful
  functor from $\MS$ to $\Top$, which is $\Po$-enriched (see
  Definition~\ref{def:monad:lift} and the subsequent remark), thus one
  may wonder if $\hat{\PM}_S$ is a lifting of a (possibly $\Po$-enriched)
  monad on $\Top$.
  This is impossible, because the robust topology $\tau_{d,R}$, and
  thus the open ball topology $\tau_{d_S}$, is not determined by
  $\tau_d$ (see Example~\ref{ex:tauR_counterexample}).
\end{remark}
We recall the definition of the powerset monad $\hat{\PM}$ on $\Set$.
\begin{definition}\label{def:PM}
The powerset monad $\hat{\PM}$ on $\Set$ is given by the triple
$(\PM,\eta,-^*)$, where
\begin{itemize}
\item $\PM(X)$ is the set of subsets of $X$
\item $\eta_X:\Set(X,\PM(X))$ is $\eta_X(x)=\{x\}$
\item if $f:\Set(X,\PM(X'))$, then $f^*:\Set(\PM(X),\PM(X'))$ is
  $f^*(A)=\bigcup\setf{f(x) \mid x\in A}$.
\end{itemize}
\end{definition}
In~\cite{DFM:ICTAC:2023},
we defined for each (not necessarily continuous) quantale $Q$ a
$\Po$-enriched
monad\footnote{In~\cite{DFM:ICTAC:2023}
this monad is denoted $\hat{\PM}_S$, but to avoid confusion we denote
it $\hat{\PM}_Q$ here.} $\hat{\PM}_Q$ on $\MS_Q$ such that
$\PM_Q(X,d,Q)=(\PM(X),d_Q,Q)$.
\begin{definition}\label{def:PM_Q}
  Given a continuous quantale $Q$,
  the $\Po$-enriched monad $\hat{\PM}_Q=(\PM_Q,\eta,-^*)$ on $\MS_Q$ is
  the unique lifting of $\hat{\PM}$ along $U:\MS_Q\to\Set$
  such that $\PM_Q(X,d,Q)=(\PM(X),d_Q,Q)$, where
  \begin{equation}
    \label{eq:def_d_Q}
    d_Q(A_1,A_2)\defeq\qM_{x_2\in A_2}\qJ_{x_1 \in A_1}d(x_1,x_2).
  \end{equation}
\end{definition}
If $Q=\RQ_+$, then the metric $d_Q$ is the Hausdorff-Smyth hemi-metric
defined in~\cite{goubault2008simulation}. As proved
in~\cite[Proposition 12]{DFM:ICTAC:2023}, the monad $\hat{\PM}_Q$
captures the robust topology when the quantale $Q$ is non-trivial and
linear (thus continuous), but for a generic continuous quantale only
$\tau_{d_S}\subseteq\tau_{d,R}$ holds.
\begin{example}
\label{example:coutnerexample_non_linear_quantale}
We show that for a finite (thus continuous) non-linear
locale $Q$ and a finite $Q$-metric space $(X,d,Q):\MS_Q$ there
is no $(\PM(X),d',Q):\MS_Q$ such that $\tau_{d'}=\tau_{d,R}$.
If $Q\in\Quan$ is affine (\ie, $\qI=\qt$) and $\qI \ll \qI$, then
for every $(X,d,Q):\MS_Q$
\begin{itemize}
\item $\tau_d$ is generated by the open balls with radius $\qI$ and is
  the Alexandrov topology for $\leq_d$
\item for every $U\subseteq X$ one has $B_R(U,\qI)=\setb{v}{\exists u\in
  U.u\leq_dv}$, therefore the robust topology $\tau_{d,R}$ is given by
  $O\in\tau_{d,R}\iff O\subseteq\PM(X)$ and $\forall U\in O.\forall
  V\subseteq B_R(U,\qI).V\in O$
\item if $\leq_d$ is the equality on $X$, then $B_R(U,\qI)=U$ and
  $O\in\tau_{d,R}\iff O\subseteq\PM(X)$ downward closed.
\end{itemize}
Let $Q$ be the locale $\Sigma^2$ (\ie,
$Q=\setf{\qb\qle\qt_0,\qt_1\qle\qt}$) and consider
\begin{itemize}
\item the sets $X=\setf{x_0,x_1,x}$ and $A=\setf{x_0,x_1}$
\item $d:X^2\to Q$ the unique symmetric separated metric such that
  $d(x_i,x)=\qt_i$ and $d(x_0,x_1)=\bot$
\item $d':\PM(X)^2\to Q$ a metric such that
  $\eta_X:(X,d,Q)\to(\PM(X),d',Q)$ is an arrow in $\MS_Q$ (\ie, a
  short map), where $\eta$ is the unit of the powerset monad on
  $\Set$, \ie, $\forall u,v:X.d(u,v)\qle d'(\setf{u},\setf{v})$.
\end{itemize}
We prove that $\tau_{d'}\neq\tau_{d,R}$ by contradiction.
If $\tau_{d'}=\tau_{d,R}$, then:
\begin{itemize}
\item $O\in\tau_{d'}\iff O\subseteq\PM(X)$ downward closed, therefore
  $U\leq_{d'} V\iff U\supseteq V$.
\item  If $i\in\setf{0,1}$, then
  $A\leq_{d'}\setf{x_i}$, \ie, $d'(A,\setf{x_i})=\qt$.
\item Therefore, $\forall i\in\setf{0,1}.
  \qt_i=d(x_i,x)\qle d'(A,\setf{x_i})\qm d'(\setf{x_i},\setf{x})\qle
  d'(A,\setf{x})$.
\end{itemize}
Hence, we get a contradiction, because $x\notin A$, but
$\qt=\qt_0\qj\qt_1\qle d'(A,\setf{x})$, which implies
$A\supseteq\setf{x}$.
  \end{example}

In order to define the $\Po$-enriched monad $\hat{\PM}_S$ on $\MS$ we
proceed as follow:
\begin{enumerate}
\item We define a monad $\hat{\CM}$ on the category $\CL$ of continuous
  lattices (see Definition~\ref{def:CL}).
\item We define a lifting $\hat{\CM}_S$ of $\hat{\CM}$ along the forgetful functor
  $\Quan\to\CL$.
\item Finally, we define $\hat{\PM}_S$ as the unique lifting of
  $\hat{\PM}$ along the forgetful functor $\MS\to\Set$ such that
  $\PM_S(X,d,Q)=(\PM(X),d_S,\CM_S(Q))$, for a suitable metric $d_S$.
\end{enumerate}

The monad $\hat{\CM}$ on the category $\CL$ of continuous lattices and Scott
continuous maps resembles the powerset monad, but the elements in
$\CM(Q)$ are Scott closed subsets of $Q$, instead of arbitrary
subsets.
There are some proof obligations in order to claim that the
definition below is a monad on $\CL$, therefore the definition is
followed by a proof of such obligations.
The category $\CL$ is $\Po$-enriched, and one can prove that
$\hat{\CM}$ is $\Po$-enriched, but this additional property is not
needed for our purposes.
\begin{definition}\label{def:CM}
The monad $\hat{\CM}$ on $\CL$ is given by the triple
$(\CM,\eta,-^*)$, where
\begin{itemize}
\item If $Q:\CL$, then $\CM(Q)$ is the continuous lattice of Scott
  closed subsets of $Q$ ordered by inclusion
\item If $Q:\CL$, then $\eta_Q:\CL(Q,\CM(Q))$ is
  $\eta_Q(q)=\cl{\{q\}}$, where $\cl{A}$ denotes the closure of the
  subset $A\subseteq Q$
\item If $g:\CL(Q,\CM(Q'))$, then $g^*:\CL(\CM(Q),\CM(Q'))$ is
  $g^*(A)=\cl{\bigcup\setf{g(q) \mid q\in A}}$.
\end{itemize}
\end{definition}
\begin{proof}
To claim that $(\CM,\eta,-^*)$ is a monad on $\CL$ we have the
following proof obligations:
\begin{enumerate}
\item The poset $\CM(Q):\Pos$ is in $\CL$.
\item The map $\eta_Q:\Set(Q,\CM(Q))$ is in $\CL(Q,\CM(Q))$.
\item The map $g^*:\Set(\CM(Q),\CM(Q'))$ is in $\CL(\CM(Q),\CM(Q'))$
  when $g:\CL(Q,\CM(Q'))$.
\item The three equations in Definition~\ref{def:po-monad} hold.
\end{enumerate}
We prove each item individually.
\begin{enumerate}
\item It is easy to see that $\CM(Q)$ is a complete lattice.  In fact,
  for any $S\subseteq\CM(Q)$ the meet $\qM S$ is the intersection
  $\bigcap S$ and the join $\qJ S$ is the closure of the union
  $\cl{\bigcup S}$.  We have to define a base.

  Recall that a subset $A\subseteq Q$ is Scott closed $\iff$ it is
  lower (\ie $A=\qLowerSet{A}$) and closed under sups of directed
  subsets (\ie, if $D\subseteq A$ is directed, then $\qJ D\in A$).
  Thus, if $A\subseteq_f Q$, then $\cl{A}=\qLowerSet{A}$.

  Therefore, if $B$ is a base for $Q$, then $\setf{\cl{A} \mid A\subseteq_f
    B}$ is a base for $\CM(Q)$.
  
\item $\eta_Q:\Set(Q,\CM(Q))$ is monotonic (because $p\qle q$ implies
  $\qLowerSet{p}\subseteq \qLowerSet{q}$) and Scott continuous,
  because if $A$ is a directed subsets of $Q$, then
  $S=\setf{\qLowerSet{q} \mid q\in A}$ is a directed subset of
  $\CM(Q)$, $\bigcup S=\qLowerSet{A}$ is a directed subset of $Q$, and
  therefore $\cl{\bigcup S} = \qLowerSet{(\qJ A)}$.

\item If $g:\CL(Q,\CM(Q'))$, then $g^*:\Set(\CM(Q),\CM(Q'))$ is monotonic
  (because $A\subseteq B$ in $\CM(Q)$ implies $g^*(A)\subseteq g^*(B)$
  in $\CM(Q')$) and join-preserving (thus Scott continuous), because
  it has a right-adjoint (in $\Pos$) given by $g^-(A')\defeq\setf{q\in
    Q \mid g(q)\subseteq A'}$.

  First, we prove that $A=g^-(A')\in\CM(Q)$.  In fact, $A$ is a lower
  subset (\ie, $p\qle q\in A$ implies $p\in A$), because
  $g(p)\subseteq g(q)$ by monotonicity of $g$, and closed under sups
  of directed subsets, because $g$ is Scott continuous.  It is obvious
  that $g^-$ is monotonic, \ie, $g^-:\Pos(\CM(Q'),\CM(Q))$.  To prove
  that $g^*\dashv g^-$ (in $\Pos$) we show that $A\subseteq
  g^-(A')\iff g^*(A)\subseteq A'$:
  \begin{eqnarray*}
    & &A\subseteq g^-(A')\\
    (\text{by definition of $g^-$}) & \iff & \forall q\in
                                              A. \ g(q)\subseteq A' \\
    & \iff & \bigcup\setf{g(q) \mid q\in A}\subseteq A' \\
    (\text{because $A'$ is closed}) & \iff & \cl{\bigcup\setf{g(q)
                                             \mid q\in A}}\subseteq A'
    \\
    (\text{by definition of $g^*$}) & \iff & g^*(A)\subseteq A'.    
  \end{eqnarray*}

\item If $f:\CL(Q,\CM(Q'))$ and $g:\CL(Q',\CM(Q''))$, then
  $f^*\circ\eta_Q=f$, $\eta_Q^*=\id_{\CM(Q)}$ and $g^*\circ
  f^*=(g^*\circ f)^*$.
  
  Since $f^*$ and $\id_{\CM(Q)}$ are join-preserving, to prove the
  last two equations it suffices to check that they hold for elements
  of the form $\eta_Q(q)=\qLowerSet{q}$ in $\CM(Q)$.
  \begin{itemize}
  \item $f^*(\qLowerSet{q})=f(q)$, because $f(p)\subseteq f(q)$ when
    $p\in\qLowerSet{q}$.
  \item
    $\eta_Q^*(\qLowerSet{q})=\eta_Q(q)= \qLowerSet{q}
    =\id_{\CM(Q)}(\qLowerSet{q})$, by replacing $f$ with $\eta_Q$ in
    the first equation.
  \item $g^*(f^*(\qLowerSet{q}))=g^*(f(q))=(g^*\circ f)^*(\qLowerSet{q})$,
    by using two instances of the first equation.
  \end{itemize}
\end{enumerate}
\end{proof}
We now define a lifting $\hat{\CM}_S$ of $\hat{\CM}$ along the
forgetful functor $U:\Quan\to\CL$ mapping a continuous quantale
$(Q,\qT)$ to the underlying continuous lattice $Q$.
Since the functor $U$ is faithful, the lifting is uniquely determined
by the action of $\hat{\CM}_S$ on the objects of $\Quan$, and the
three equations in Definition~\ref{def:po-monad} are automatically
satisfied (thus they are not part of the proof obligations).
\begin{definition}\label{def:CM_S}
  The monad $\hat{\CM}_S=(\CM_S,\eta,-^*)$ on $\Quan$ is the unique
  lifting of $\hat{\CM}$ along $U:\Quan\to\CL$ such that
  $\CM_S(Q,\qT)=(\CM(Q),\qT_\CM)$, where $A_1\qT_\CM A_2\defeq
  \cl{\setb{q_1\qT q_2}{q_1\in A_1\land q_2\in A_2}}$.
\end{definition}
\begin{proof}
To claim that $(\CM_S,\eta,-^*)$ is a monad on $\Quan$ we have the
following proof obligations:
\begin{enumerate}
\item $(\CM(Q),\qT_\CM)$ is in $\Quan$.
\item $\eta_Q:\CL(Q,\CM(Q))$ is in $\Quan((Q,\qT),(\CM(Q),\qT_\CM))$.
\item $g^*:\CL(\CM(Q),\CM(Q'))$ is in
  $\Quan((\CM(Q),\qT_\CM),(\CM(Q'),\qT_\CM'))$ when
  $g:\Quan((Q,\qT),(\CM(Q'),\qT_\CM'))$.
\end{enumerate}
First we related the continuous lattice $\CM(Q)$ to the algebraic
lattice $\DM(Q)$ defined in Proposition~\ref{prop-D-construction}:

\begin{itemize}
\item There is an adjunction
    \begin{tikzcd}[column sep = large]
    \CM(Q) \arrow[r, hookrightarrow , yshift =
    -1.3ex] & \DM(Q)  \arrow[l, dashed , "\bot", yshift = 1.1ex]
  \end{tikzcd}in $\Pos$, where the left-adjoint maps $A\in \DM(Q)$ to
  $\cl{A}\in\CM(Q)$, moreover for any $S\subseteq \DM(Q)$ the join
  $\qJ\setf{\cl{A} \mid A\in S}$ in $\CM(Q)$ is $\cl{\bigcup S}$.
\item 
  Proposition~\ref{prop-D-alg} implies that $(\DM(Q),\hat{\qT})$ is a
  quantale when $(Q,\qT)$ is a quantale, and for any
  $A_1,A_2\in\DM(Q)$ the tensor $\cl{A_1}\qT_\CM\cl{A_2}$ is
  $\cl{A_1\hat{\qT}A_2}$ , because
  $\cl{A_1}\hat{\qT}\cl{A_2}\subseteq\cl{A_1\hat{\qT}A_2}$ by
  continuity of $\qT$.
\end{itemize}
Now we can prove the three items.
\begin{enumerate}
\item Using the relations between $\CM(Q)$ and $\DM(Q)$, we prove
  that $\qT_\CM$ distributes over joins, \ie,\\
  $(\qJ\setf{\cl{A} \mid A\in S_1})\qT_\CM(\qJ\setf{\cl{A} \mid A\in S_2})=
  \qJ\setf{\cl{A}_1\qT_\CM\cl{A}_2 \mid A_1\in S_1\land A_2\in S_2}$
  when $S_1,S_2\subseteq\DM(Q)$:

  \begin{eqnarray*}
    & & (\qJ\setf{\cl{A} \mid A\in S_1})\qT_\CM(\qJ\setf{\cl{A} \mid A\in
        S_2}) \\
     (\text{because $\cl{\bigcup S}=\cl{\bigcup\setf{\cl{A} \mid A\in
     S}}$}) & =&  \cl{\bigcup S_1}\qT_\CM\cl{\bigcup S_2} \\
    (\text{by the relation
    between $\qT_\CM$ and $\hat{\qT}$}) & = &  \cl{(\bigcup
                                              S_1)\hat{\qT}(\bigcup
                                              S_2)} \\
    (\text{by distributivity
                                              for $\hat{\qT}$}) & =
      &\cl{\bigcup\setf{A_1\hat{\qT}A_2 \mid A_1\in S_1\land A_2\in
        S_2}} \\
(\text{because $\cl{\bigcup S}=\cl{\bigcup\setf{\cl{A} \mid A\in
    S}}$})    & = & \cl{\bigcup\setf{\cl{A_1\hat{\qT}A_2} \mid A_1\in S_1\land A_2\in
                    S_2}}\\
    & = & \qJ\setf{\cl{A}_1\qT_\CM\cl{A}_2 \mid A_1\in S_1\land A_2\in
    S_2}.    
  \end{eqnarray*}

  Moreover, the unit $\qI_\CM$ and $\hat{\qI}$ are the same, namely
  $\qLowerSet{\qI} = \eta_Q(\qI)$.

\item By Proposition~\ref{prop-D-construction} $\eta_Q$ is
  strict-monoidal, thus lax-unital.

\item The map $g^*$ is lax-unital when
  $\eta_{Q'}(\qI')\subseteq g^*(\eta_Q(\qI))$.  The proof in
  Definition~\ref{def:CM} shows that $g^*\circ\eta_Q=g$, therefore
  $\eta_{Q'}(\qI')\subseteq g(\qI)=g^*(\eta_Q(\qI))$ when $g$ is
  lax-unital.
\end{enumerate}
\end{proof}
\begin{remark}
  The monad $\hat{\CM}_S$ restricts to several full sub-categories of
  $\Quan$, this is because the function $\CM_S$ on the objects of
  $\Quan$ has the following properties:
  \begin{itemize}
  \item it maps finite quantales to finite quantales
  \item it maps linear quantales to linear quantales, and in this cases
    $\CM(Q)$ is isomorphic to the lifting $Q_\bot$
  \item  it maps $\omega$-continuous quantales to $\omega$-continuous quantales
  \item it maps affine quantales to affine quantales
  \item it maps commutative quantales to commutative quantales
  \item it maps locales to locales.
  \end{itemize}    
\end{remark}

\begin{definition}\label{def:PM_S}
  If $\A$ is a sub-category of $\Quan$ such that the monad
  $\hat{\CM}_S$ on $\Quan$ restricts to $\A$, then the $\Po$-enriched
  monad $\hat{\PM}_S=(\PM_S,\eta,-^*)$ on $\MS_\A$ is the unique
  lifting of $\hat{\PM}$ along $U:\MS_\A\to\Set$ such that
  $\PM_S(X,d,Q)=(\PM(X),d_S,\CM_S(Q))$, where
  \begin{equation}
    \label{eq:def_d_S}
    d_S(A_1,A_2)\defeq\bigcap_{x_2\in A_2}\cl{\setb{d(x_1,x_2)}{x_1
        \in A_1}}\in\CM_S(Q).
  \end{equation}
\end{definition}
\begin{proof}
To claim that $(\PM_S,\eta,-^*)$ is a $\Po$-enriched monad on $\MS$
(and consequently on $\MS_\A$ when $\hat{\CM}_S$ restricts to $\A$) we
have the following proof obligations:
\begin{enumerate}
\item $d_S:\PM(X)^2\to\CM_S(Q)$ is a metric.
\item $\eta_X:\Set(X,\PM(X))$ is in $\MS((X,d,Q),(\PM(X),d_S,\CM(Q)))$.
\item $f^*:\Set(\PM(X),\PM(X'))$ is in
  $\MS((\PM(X),d_S,\CM_S(Q)),(\PM(X'),d_S',\CM_S(Q')))$ when\\
  $f:\MS((X,d,Q),(\PM(X'),d_S',\CM_S(Q')))$.
\item $-^*$ is monotonic with respect to hom-preorders of $\MS$.
\end{enumerate}
We prove each item individually.
\begin{enumerate}
\item $d_S$ is an instance of the metric $d_Q$ (see
  Definition~\ref{def:PM_Q}) with $Q$ replaced by $\CM_S(Q)$ and $d$
  by $\eta_Q\circ d$:
  \begin{itemize}
  \item $d'\defeq\eta_Q\circ d:X^2\to\CM_S(Q)$ is a metric, because
    $\eta_Q$ is strict-monoidal (thus lax-monoidal)
  \item $d_S=d'_{\CM_S(Q)}$, because $d'(x,y)=\qLowerSet{d(x,y)}$ and
    by expanding the definitions $\qM$ and $\qJ$ in $\CM(Q)$.
\end{itemize}
\item 
  $\eta_Q:\Quan(Q,\CM_S(Q))$ of Definition~\ref{def:CM_S} is a
  realizer of $\eta_X:\MS((X,d,Q),(\PM(X),d_S,\CM(Q)))$, because\\
  $\eta_Q(d(x,y))=\qLowerSet{d(x,y)}=\cl{\setf{d(x,y)}}=d_S(\setf{x},\setf{y})=d_S(\eta_X(x),\eta_X(y))$
  in $\CM_S(Q)$.
\item Given a realizer $g:\Quan(Q,\CM_S(Q'))$ of
  $f:\MS((X,d,Q),(\PM(X'),d_S',\CM_S(Q')))$ we prove that a realizer of
  $f^*:\MS((\PM(X),d_S,\CM_S(Q)),(\PM(X'),d_S',\CM_S(Q')))$ is given by
  $g^*:\Quan(\CM_S(Q),\CM_S(Q'))$ of Definition~\ref{def:CM_S}, \ie,
  $g^*(d_S(A_1,A_2))\subseteq d'_S(f^*(A_1),f^*(A_2))$.
  \begin{eqnarray*}
    & & g^*(d_S(A_1,A_2)) \\
(\text{by definition of $d_S$})    & =& g^*(\bigcap_{x_2\in
                                         A_2}\qJ_{x_1\in
                                         A_1}\eta_Q(d(x_1,x_2))) \\
(\text{by monotonicity of $g^*$})    & \subseteq & \bigcap_{x_2\in
                                                   A_2}g^*(\qJ_{x_1\in
                                                   A_1}\eta_Q(d(x_1,x_2)))
    \\
(\text{by $g^*$ being join-preserving})    & = & \bigcap_{x_2\in
                                           A_2}\qJ_{x_1\in
                                           A_1}g^*(\eta_Q(d(x_1,x_2)))\\
(\text{by the monad law $g^*\circ\eta_q=g$})    & = &  \bigcap_{x_2\in
                                                      A_2}\qJ_{x_1\in
                                                      A_1}g(d(x_1,x_2))
    \\
(\text{since $g$ is a realizer of $f$})    & \subseteq &
                                                         \bigcap_{x_2\in
                                                         A_2}\qJ_{x_1\in
                                                         A_1}
                                                         d'_S(f(x_1),f(x_2))
    \\
(\text{by definition of $d'_S$})    & = & \bigcap_{x_2\in A_2}\qJ_{x_1\in A_1} \bigcap_{x'_2\in
                                          f(x_2)}\qJ_{x'_1\in f(x_1)}\eta_{Q'}(d'(x_1,x_2)) \\
(\text{$\qJ_i\qM_j q_{i,j}\qle \qM_j\qJ_i q_{i,j}$ holds in any complete
    lattice})    & \subseteq & \bigcap_{x_2\in A_2}\bigcap_{x'_2\in f(x_2)}\qJ_{x_1\in
                               A_1}\qJ_{x'_1\in f(x_1)}\eta_{Q'}(d'(x_1,x_2)) \\
(\text{by definition of
    $f^*(A)$})    & =& \bigcap_{x'_2\in f^*(A_2)}\qJ_{x'_1\in
                       f^*(A_1)}\eta_{Q'}(d'(x_1,x_2))\\
(\text{by definition of $d'_S$})    & = & d'_S(f^*(A_1),f^*(A_2)).
  \end{eqnarray*}

\item
  If $f_1\leq f_2:\MS((X,d,Q),(\PM(X'),d'_S,\CM_S(Q')))$, \ie,
  $\forall x:X.\qI'\in d'_S(f_1(x),f_2(x))$, then we must show that
  $\forall A:\PM(X).\qI'\in d'_S(f_1^*(A),f_2^*(A))$.

  Since $\qI'\in d'_S(A'_1,A'_2)$ is equivalent to $\forall x'_2\in
  A'_2.\exists x'_1\in A'_1.\qI'\qle' d'(x'_1,x'_2)$, the assumption
  is equivalent to $\forall x\in X.\forall x'_2\in f_2(x).\exists
  x'_1\in f_1(x).\qI'\qle' d'(x'_1,x'_2)$, and we have:

  \begin{eqnarray*}
    & & \qI'\in d'_S(f_1^*(A),f_2^*(A))\\
 & \iff &  \forall x'_2\in f_2^*(A). \ \exists x'_1\in f_1^*(A). \ \qI'\qle'
                                         d'(x'_1,x'_2) \\
(\text{by definition of $f_2^*(A)$})    & \iff & \forall x\in A. \ \forall x'_2\in f_2(x). \ \exists x'_1\in
    f_1^*(A). \ \qI'\qle' d'(x'_1,x'_2),     
  \end{eqnarray*}
  which follows from the assumption by taking $x'_1\in f_1(x)$ such
  that $\qI'\qle' d'(x'_1,x'_2)$.
\end{enumerate}
\end{proof}

Finally, we prove that the metric $d_S$ in Definition~\ref{def:PM_S}
captures the robust topology for the metric~$d$.
\begin{theorem}
  \label{theorem:robust_top_induced_by_Hausdorff}
  If $(X,d,Q):\MS$, then $\tau_{d,R}=\tau_{d_S}$, where
  $(\PM(X),d_S,\CM_S(Q))=\PM_S(X,d,Q)$.
\end{theorem}
\begin{proof} 
  We prove that each topology is included in the other.
\begin{description}
\item[$\tau_{d_S}\subseteq\tau_{d,R}$.] It suffices to
  prove that, for any $A\in\PM(X)$ and $\Delta\subseteq_f Q$ such that
  $\qLowerSet{\Delta}\ll\qLowerSet{\qI}$ in $\CM_S(Q)$ (or equivalently,
  $\qJ\Delta\ll\qI$ in $Q$) the open ball $B(A,\qLowerSet{\Delta})$ in
  $(\PM(X),d_S,\CM_S(Q))$ is in $\tau_{d,R}$, \ie,
  \begin{enumerate}
  \item $B(A,\qLowerSet{\Delta})$ downward closed and
  \item $A'\in B(A,\qLowerSet{\Delta}) \implies 
    \exists\delta\ll\qI. B_R(A',\delta)\in B(A,\qLowerSet{\Delta})$.
  \end{enumerate}
  The first property is immediate.
  For the second, take $A'\in B(A,\qLowerSet{\Delta})$, \ie,
  $\qLowerSet{\Delta}\ll d_S(A,A')$.  By Lemma~\ref{lemma:inter:cq}, we
  get $\qLowerSet{\Delta}\ll d_S(A,A')\qT_\CM\qLowerSet{\Delta'}$ for some
  $\Delta'\subseteq_f Q$ such that $\qLowerSet{\Delta'}\ll
  \qLowerSet{\qI}$.

  Let $\delta \defeq \qJ \Delta'$, then $\qLowerSet{\delta}\subseteq
  d_S(A',B_R(A',\delta))$, or equivalently $\forall y\in
  B_R(A',\delta).\exists x\in A'.\delta \qle d(x,y)$, because
  $B_R(A',\delta)=\setb{y\in X}{\exists x\in A'.\delta\ll d(x,y)}$.

  We can now prove $B_R(A',\delta)\in B(A,\qLowerSet{\Delta})$, \ie,
  $\qLowerSet{\Delta}\subseteq d_S(A,B_R(A',\delta))$:

  \begin{eqnarray*}
    & & \qLowerSet{\Delta} \\
(\text{by definition of $\Delta'$})    & \ll &
                                               d_S(A,A')\qT_\CM\qLowerSet{\Delta'}
    \\
(\text{because
    $\delta=\qJ\Delta'$, thus $\Delta'\subseteq\qLowerSet{\delta}$})
    & \subseteq & d_S(A,A')\qT_\CM\qLowerSet{\delta} \\
(\text{because
    $\qLowerSet{\delta}\subseteq d_S(A',B_R(A',\delta))$})    &
                                                                \subseteq
      & d_S(A,A') \qT_\CM d_S(A',B_R(A',\delta)) \\
(\text{by the
    triangle inequality})    & \subseteq & d_S(A,B_R(A',\delta)).
  \end{eqnarray*}

\item[($\tau_{d,R} \subseteq \tau_{d_S}$).] Let $U\in\tau_{d,R}$, we
  prove that $\forall A\in U. \exists \delta'\ll\qI.
  B(A,\qLowerSet{\delta'})\subseteq U$, which implies $U\in\tau_{d_S}$.

  If $A\in U\in\tau_{d,R}$, then there exists $\delta\ll\qI$ such that
  $B_R(A,\delta)\in U$, and by Lemma~\ref{lemma:inter:cl}, there
  exists $\delta'\in Q$ such that $\delta\ll\delta'\ll\qI$.

  If $A'\in B(A,\qLowerSet{\delta'})$, then $\delta'\in d_S(A,A')$, or
  equivalently $\forall y\in A'.\exists x\in A.\delta'\qle d(x,y)$,
  which implies $\forall y\in A'.\exists x\in A.\delta \ll d(x,y)$,
  \ie, $A'\subseteq B_R(A,\delta)$.  Since $B_R(A,\delta)\in U$ and
  $U\in\tau_{d,R}$ is downward closed, we conclude $A'\in U$.
\end{description}
\end{proof}
We conclude by relating $\PM_Q(X,d,Q)$ and $\PM_S(X,d,Q)$ (see
Definitions~\ref{def:PM_Q}~and~\ref{def:PM_S}, respectively).
\begin{remark}
  For any $Q\in\Quan$ the arrow $\eta_Q:\Quan(Q,\CM_S(Q))$ has a
  left-adjoint, namely $\qJ:\Quan(\CM_S(Q),Q)$, which is
  strict-monoidal and join-preserving.  Since $\qJ d_S(A_1,A_2)\qle
  d_Q(A_1,A_2)$ for every $A_1,A_2\subseteq X$, we conclude that
  $\id_{\PM(X)}:\MS(\PM_S(X,d,Q),\PM_Q(X,d,Q))$, more precisely the arrow
  is in $\MS_\A$, where $\A$ is the sub-category of $\Quan$ whose
  arrows are strict-monoidal and join-preserving.

  If $Q$ is linear, then $\CM_S(Q)$ is isomorphic to $Q_\bot$, namely
  every $A\in\CM_S(Q)$ is either $\qLowerSet{q}$ for a unique $q\in Q$
  or is the empty set $\emptyset$.  If $Q$ is also non-trivial, then
  $\qJ:\Quan(\CM_S(Q),Q)$ has a (unique) left-adjoint
  $f:\Quan(Q,\CM_S(Q))$, which maps $\qb$ to $\emptyset$ and
  $q\sqsupset\qb$ to $\qLowerSet{q}$, where the assumption $Q$
  non-trivial is needed to ensure that $f$ is lax-unital.  In
  conclusion, if $Q$ is non-trivial and linear, then
  $\id_{\PM(X)}:\MS(\PM_S(X,d,Q),\PM_Q(X,d,Q))$ is an isomorphism,
  because $f$ is a realizer for the inverse, \ie, $f(d_Q(A_1,A_2))
  \subseteq d_S(A_1,A_2)$ for every $A_1,A_2\subseteq X$.
\end{remark}

\section{Concluding Remarks}

We have presented a categorical framework for robustness analysis of
systems modeled by quantale-valued metric spaces. The framework is
constructed over the class of continuous quantales.
If, in addition, the lattices are
effectively given, then we have a foundation for computable analysis
on quantale-valued metric spaces. We have not discussed computability
in the current article, and leave a detailed study of it for future
work.

The extensive metrization result of
Theorem~\ref{thm:extensive_metrizability} shows that every topology is
induced by a quantale-valued metric. As such, our framework enables
quantitative treatment of imprecision and robustness for any
topological space. A direction for future work involves investigating
how topological properties ({\eg}, sobriety, coherence, etc.) of the
topological space $(X,\tau_d)$ corresponding to a metric space
$(X,d,Q)\in\MS$ are related to properties of the quantale $Q$ and/or
of the $Q$-metric $d$~\cite{Zhang_Zhang:Sober_Quantale:2022}.

A key result is the construction of a preorder-enriched monad on the
category $\MS$ of quantale-valued metric spaces
(Definition~\ref{def:PM_S}) which captures the robust topology
(Theorem~\ref{theorem:robust_top_induced_by_Hausdorff}). For this, we
had to allow quantales to vary across objects in the category, in
contrast with~\cite{DFM:ICTAC:2023}, where the quantale $Q$ was fixed
for all the objects in the category $\MS_Q$ (see,
\cite[Deﬁnition~15]{DFM:ICTAC:2023}), and the robust topology could be
captured only when $Q$ was a linear quantale~\cite[Propositions~11
  and 12]{DFM:ICTAC:2023}. We provided a further justification of the
need to allow quantales to vary in
Example~\ref{example:coutnerexample_non_linear_quantale}.

In Example~\ref{ex:tauR_counterexample}, we showed that, in general,
the robust topology is not determined by the open ball topology. To
address this problem, we focused on categories of quantale-valued
metric spaces with uniformly continuous morphisms. Another approach,
which will be studied in future work, is to consider uniform spaces~\cite{James87}.



\begin{thebibliography}{GHK{\etalchar{+}}03}

\bibitem[AFMS21]{Adamek2021}
Jiří Adámek, Chase Ford, Stefan Milius, and Lutz Schröder.
\newblock Finitary monads on the category of posets.
\newblock {\em Mathematical Structures in Computer Science}, 31(7):799–821,
  2021.

\bibitem[AJ94]{AbramskyJung94-DT}
Samson Abramsky and Achim Jung.
\newblock Domain theory.
\newblock In S.~Abramsky, D.~M. Gabbay, and T.~S.~E. Maibaum, editors, {\em
  Handbook of Logic in Computer Science}, volume~3, pages 1--168. Clarendon
  Press, Oxford, 1994.

\bibitem[AV93]{AbramskyV93}
Samson Abramsky and Steven Vickers.
\newblock Quantales, observational logic and process semantics.
\newblock {\em Mathematical Structures in Computer Science}, 3(2):161--227,
  1993.

\bibitem[BKP18]{BonchiKP18}
Filippo Bonchi, Barbara K{\"{o}}nig, and Daniela Petrisan.
\newblock Up-to techniques for behavioural metrics via fibrations.
\newblock In Sven Schewe and Lijun Zhang, editors, {\em 29th International
  Conference on Concurrency Theory, {CONCUR} 2018}, volume 118 of {\em LIPIcs},
  pages 17:1--17:17. Schloss Dagstuhl - Leibniz-Zentrum f{\"{u}}r Informatik,
  2018.

\bibitem[Bor94]{borceux1994handbook}
Francis Borceux.
\newblock {\em Handbook of categorical algebra: volume 1, Basic category
  theory}, volume~1.
\newblock Cambridge University Press, 1994.

\bibitem[CC92]{cousot1992abstract}
Patrick Cousot and Radhia Cousot.
\newblock Abstract interpretation frameworks.
\newblock {\em Journal of logic and computation}, 2(4):511--547, 1992.

\bibitem[CW21]{CookW21}
Derek~S. Cook and Ittay Weiss.
\newblock The topology of a quantale valued metric space.
\newblock {\em Fuzzy Sets and Systems}, 406:42--57, 2021.

\bibitem[CW22]{Cook_Weiss:quantales_Lip:2022}
Derek~S. Cook and Ittay Weiss.
\newblock Diagrams of quantales and lipschitz norms.
\newblock {\em Fuzzy Sets and Systems}, 444:79--102, 2022.

\bibitem[DFM23]{DFM:ICTAC:2023}
Francesco Dagnino, Amin Farjudian, and Eugenio Moggi.
\newblock Robustness in metric spaces over continuous quantales and the
  {Hausdorff-Smyth} monad.
\newblock In Erika {\'A}brah{\'a}m, Clemens Dubslaff, and Silvia Lizeth~Tapia
  Tarifa, editors, {\em Theoretical Aspects of Computing -- ICTAC 2023}, volume
  14446 of {\em Lecture Notes in Computer Science}, pages 313--331, Cham, 2023.
  Springer Nature Switzerland.

\bibitem[DGY19]{DalLagoGY19}
Ugo {Dal Lago}, Francesco Gavazzo, and Akira Yoshimizu.
\newblock Differential logical relations, part {I:} the simply-typed case.
\newblock In Christel Baier, Ioannis Chatzigiannakis, Paola Flocchini, and
  Stefano Leonardi, editors, {\em 46th International Conference on Automata,
  Languages and Programming, {ICALP} 2018}, volume 132 of {\em LIPIcs}, pages
  111:1--111:14. Schloss Dagstuhl - Leibniz-Zentrum f{\"{u}}r Informatik, 2019.

\bibitem[DN23]{DahlqvistN23}
Fredrik Dahlqvist and Renato Neves.
\newblock The syntactic side of autonomous categories enriched over generalised
  metric spaces.
\newblock {\em Log. Methods Comput. Sci.}, 19(4), 2023.

\bibitem[EKL25]{Eklund_Kortelainen_Lofstrand:Quantales_Higher_Types:2025}
Patrik Eklund, Jari Kortelainen, and Magnus L{\"o}fstrand.
\newblock Quantales for fuzzy sets and relations of higher types.
\newblock {\em Mathematics}, 13(13), 2025.

\bibitem[FK97]{FlaggK97}
Bob Flagg and Ralph Kopperman.
\newblock Continuity spaces: Reconciling domains and metric spaces.
\newblock {\em Theoretical Computer Science}, 177(1):111--138, 1997.

\bibitem[Fla97]{Flagg:Quantales_continuity_spaces:1997}
R.C. Flagg.
\newblock Quantales and continuity spaces.
\newblock {\em Algebra Universalis}, 37(3):257--276, 1997.

\bibitem[FM23]{Farjudian_Moggi:Robustness_Scott_Continuity_Computability:2023}
Amin Farjudian and Eugenio Moggi.
\newblock Robustness, {Scott} continuity, and computability.
\newblock {\em Mathematical Structures in Computer Science}, page 1–37, 2023.

\bibitem[Gav18]{Gavazzo18}
Francesco Gavazzo.
\newblock Quantitative behavioural reasoning for higher-order effectful
  programs: Applicative distances.
\newblock In Anuj Dawar and Erich Gr{\"{a}}del, editors, {\em Proceedings of
  the 33rd Annual {ACM}/{IEEE} Symposium on Logic in Computer Science, {LICS}
  2018}, pages 452--461. {ACM}, 2018.

\bibitem[GF23]{GavazzoF23}
Francesco Gavazzo and Cecilia~Di Florio.
\newblock Elements of quantitative rewriting.
\newblock {\em Proceedings of the ACM on Programming Languages},
  7({POPL}):1832--1863, 2023.

\bibitem[GHK{\etalchar{+}}03]{Gierz-ContinuousLattices-2003}
Gerhard Gierz, Karl~Heinrich Hofmann, Klaus Keimel, Jimmie~D. Lawson,
  Michael~W. Mislove, and Dana~S. Scott.
\newblock {\em Continuous Lattices and Domains}, volume~93 of {\em Encycloedia
  of Mathematics and its Applications}.
\newblock Cambridge University Press, 2003.

\bibitem[GHK17]{GutierrezGarcia_Hohle_Kubiak:Tensor_Quantales:2017}
Javier {Guti{\'e}rrez Garc{\'i}a}, Ulrich H{\"o}hle, and Tomasz Kubiak.
\newblock Tensor products of complete lattices and their application in
  constructing quantales.
\newblock {\em Fuzzy Sets and Systems}, 313:43--60, 2017.

\bibitem[GL08]{goubault2008simulation}
Jean Goubault-Larrecq.
\newblock Simulation hemi-metrics between infinite-state stochastic games.
\newblock {\em Lecture Notes in Computer Science}, 4962:50--65, 2008.

\bibitem[GL13]{Goubault-Larrecq:Non_Hausdorff_topology:2013}
Jean Goubault-Larrecq.
\newblock {\em Non-Hausdorff topology and domain theory}.
\newblock Cambridge University Press, 2013.

\bibitem[HK11]{Hohle_Kubiak:non_commutative_quantale:2011}
Ulrich H{\"o}hle and Tomasz Kubiak.
\newblock A non-commutative and non-idempotent theory of quantale sets.
\newblock {\em Fuzzy Sets and Systems}, 166(1):1--43, 2011.

\bibitem[HR13]{Hofmann_Reis:Prob_Met_Spaces:2013}
Dirk Hofmann and Carla~David Reis.
\newblock Probabilistic metric spaces as enriched categories.
\newblock {\em Fuzzy Sets and Systems}, 210:1--21, 2013.

\bibitem[HST14]{HofmannST14}
Dirk Hofmann, Gavin~J Seal, and Walter Tholen.
\newblock {\em Monoidal Topology: A Categorical Approach to Order, Metric, and
  Topology}, volume 153.
\newblock Cambridge University Press, 2014.

\bibitem[Jam87]{James87}
Ioan~M. James.
\newblock {\em Uniform Spaces}, pages 85--100.
\newblock Springer New York, New York, NY, 1987.

\bibitem[JM10]{JMoggi2010monad}
Mauro Jaskelioff and Eugenio Moggi.
\newblock Monad transformers as monoid transformers.
\newblock {\em Theoretical computer science}, 411(51-52):4441--4466, 2010.

\bibitem[Joh86]{Johnstone86}
Peter~T. Johnstone.
\newblock {\em Stone spaces}, volume~3 of {\em Cambridge Studies in Advanced
  Mathematics}.
\newblock Cambridge University Press, Cambridge, 1986.

\bibitem[Kel82]{kelly1982basic}
Max Kelly.
\newblock {\em Basic concepts of enriched category theory}, volume~64.
\newblock CUP Archive, 1982.
\newblock Reprints in Theory and Applications of Categories, No. 10, (2005).

\bibitem[Kop88]{Kopperman:All_topologies_metric:1988}
Ralph Kopperman.
\newblock All topologies come from generalized metrics.
\newblock {\em The American Mathematical Monthly}, 95(2):89--97, 1988.

\bibitem[KS74]{KellyStreet74}
G~Kelly and Ross Street.
\newblock Review of the elements of 2-categories.
\newblock In {\em Category Seminar}, pages 75--103. Springer, 1974.

\bibitem[Law73]{lawvere1973metric}
F.~William Lawvere.
\newblock Metric spaces, generalized logic, and closed categories.
\newblock {\em Rendiconti del seminario mat{\'e}matico e fisico di Milano},
  43:135--166, 1973.
\newblock Reprints in Theory and Applications of Categories, No. 1, 1-37
  (2002).

\bibitem[MFDT18]{Moggi_Farjudian_Duracz_Taha:Reachability_Hybrid:2018}
Eugenio Moggi, Amin Farjudian, Adam Duracz, and Walid Taha.
\newblock Safe \& robust reachability analysis of hybrid systems.
\newblock {\em Theor. Comput. Sci.}, 747:75--99, 2018.

\bibitem[MFT19]{MoggiFT-ICTCS-2019}
Eugenio Moggi, Amin Farjudian, and Walid Taha.
\newblock System analysis and robustness.
\newblock In Alessandra Cherubini, Nicoletta Sabadini, and Simone Tini,
  editors, {\em Proceedings of the 20th Italian Conference on Theoretical
  Computer Science, {ICTCS} 2019, Como, Italy, September 9-11, 2019}, volume
  2504 of {\em {CEUR} Workshop Proceedings}, pages 1--7. CEUR-WS.org, 2019.

\bibitem[MM76]{Manes1976}
Ernest~G Manes and Ernest~G Manes.
\newblock Algebraic theories in a category.
\newblock {\em Algebraic Theories}, pages 161--279, 1976.

\bibitem[Mog91]{Moggi:notions_monads:1991}
Eugenio Moggi.
\newblock Notions of computation and monads.
\newblock {\em Information and computation}, 93(1):55--92, 1991.

\bibitem[Mul86]{Mulvey:Second_topology_conference:1986}
Christopher~J. Mulvey.
\newblock Second topology conference ({Taormina}, 1984).
\newblock {\em Rend. Circ. Mat. Palermo (2) Suppl}, 12:99--104, 1986.

\bibitem[NR88]{Niefield_Rosenthal:quantales:1988}
Susan~B. Niefield and Kimmo~I. Rosenthal.
\newblock Constructing locales from quantales.
\newblock {\em Mathematical Proceedings of the Cambridge Philosophical
  Society}, 104(2):215--234, 1988.

\bibitem[Pis21]{Pistone21}
Paolo Pistone.
\newblock On generalized metric spaces for the simply typed lambda-calculus.
\newblock In {\em 36th Annual {ACM/IEEE} Symposium on Logic in Computer
  Science, {LICS} 2021}, pages 1--14. {IEEE}, 2021.

\bibitem[SKDH21]{SprungerKDH21}
David Sprunger, Shin{-}ya Katsumata, J{\'{e}}r{\'{e}}my Dubut, and Ichiro
  Hasuo.
\newblock Fibrational bisimulations and quantitative reasoning: Extended
  version.
\newblock {\em Journal of Logic and Computation}, 31(6):1526--1559, 2021.

\bibitem[Smy77]{smyth1977effectively}
Michael~B. Smyth.
\newblock Effectively given domains.
\newblock {\em Theor. Comput. Sci.}, 5(3):257--274, 1977.

\bibitem[TLZ14]{Tao_Lai_Zhang:Quantale_valued_preorders:2014}
Yuanye Tao, Hongliang Lai, and Dexue Zhang.
\newblock Quantale-valued preorders: Globalization and cocompleteness.
\newblock {\em Fuzzy Sets and Systems}, 256:236--251, 2014.

\bibitem[ZZ22]{Zhang_Zhang:Sober_Quantale:2022}
Dexue Zhang and Gao Zhang.
\newblock Sober topological spaces valued in a quantale.
\newblock {\em Fuzzy Sets and Systems}, 444:30--48, 2022.

\end{thebibliography}

\newcommand{\etalchar}[1]{$^{#1}$}

\end{document}